\documentclass{aastex631}

\usepackage[utf8]{inputenc}
\hypersetup{breaklinks}
\usepackage{comment}

\newcommand{\grp}{\ensuremath{(G - G_{\rm RP})}}
\newcommand{\gbr}{\ensuremath{(G_{\rm BP} - G_{\rm RP})}}
\newcommand{\ha}{H$\alpha$}

\newcommand\recovrate{81.4\%}

\newcommand\withtess{313}
\newcommand\rotobs{255}
\newcommand\fulltuchor{382}

\date{\today}

\begin{document}

\title{\sc{Examining the rotation period distribution of the 40 Myr Tucana-Horologium Association with TESS}}
\shorttitle{\sc{Tucana-Horologium and TESS}}
\shortauthors{Popinchalk et al.}

\newcommand{\amnh}{Department of Astrophysics, American Museum of Natural History, Central Park West at 79th Street, New York, NY 10034, USA}
\newcommand{\cunygrad}{Physics, The Graduate Center, City University of New York, New York, NY 10016, USA}
\newcommand{\hunter}{Department of Physics and Astronomy, Hunter College, City University of New York, 695 Park Avenue, New York, NY 10065, USA}
\newcommand{\columbia}{Department of Astronomy, Columbia University, 550 West 120th Street, New York, NY, USA}

\correspondingauthor{Mark Popinchalk}
\author[0000-0001-9482-7794]{Mark Popinchalk}
\affiliation{\amnh}
\affiliation{\cunygrad}
\affiliation{\hunter}
\email{popinchalkmark@gmail.com}

\author[0000-0001-6251-0573]{Jacqueline K. Faherty}
\affiliation{\amnh}

\author[0000-0002-2792-134X]{Jason L. Curtis}
\affiliation{\columbia}

\author[0000-0002-2592-9612]{Jonathan Gagn\'e}
\affiliation{Plan\'etarium Rio Tinto Alcan, Espace pour la Vie, 4801 av. Pierre-de Coubertin, Montr\'eal, Qu\'ebec, Canada}
\affiliation{Institute for Research on Exoplanets, Universit\'e de Montr\'eal, D\'epartement de Physique, C.P.~6128 Succ. Centre-ville, Montr\'eal, QC H3C~3J7, Canada}

\author[0000-0001-8170-7072]{Daniella C. Bardalez Gagliuffi}
\affiliation{Department of Physics \& Astronomy, Amherst College, 25 East Drive, Amherst, MA 01003, USA}
\affiliation{\amnh}

\author[0000-0003-0489-1528]{Johanna M.~Vos}
\affiliation{\amnh}

\author{Andrew Ayala}
\affiliation{\amnh}

\author{Lisseth Gonzales}
\affiliation{\amnh}

\author[0000-0003-2102-3159]{Rocio Kiman}
\affiliation{Kavli Institute for Theoretical Physics, University of California, Santa Barbara, CA 93106, USA}

\begin{abstract}
The Tucana-Horologium Association (Tuc-Hor) is a 40 Myr old moving group in the southern sky. In this work, we measure the rotation periods of \withtess{} Tuc-Hor objects with TESS light curves derived from TESS full frame images and membership lists driven by Gaia EDR3 kinematics and known youth indicators. We recover a period for \recovrate{} of the sample and report \rotobs{} rotaion periods for Tuc-Hor objects. From these objects we identify 11 candidate binaries based on multiple periodic signals or outlier Gaia DR2 and EDR3 re-normalised unit weight error (RUWE) values. We also identify three new complex rotators (rapidly rotating M dwarf objects with intricate light curve morphology) within our sample. Along with the six previously known complex rotators that belong to Tuc-Hor, we compare their light curve morphology between TESS Cycle 1 and Cycle 3 and find they change substantially. Furthermore, we provide context for the entire Tuc-Hor rotation sample by describing the rotation period distributions alongside other youth indicators such as \ha{} and Li equivalent width, as well as near ultra-violet and X ray flux. We find that measuring rotation periods with TESS to be a fast and effective means to confirm members in young moving groups.
\end{abstract}


\section{Introduction}

The identification and characterization of young stellar associations is experiencing a renaissance with the ESA Gaia Mission as its harbinger \citep{gaia_edr3}. Precision astrometry has revealed new co-moving collections of stars due to their spatial and velocity structure \citep[e.g.][]{Oh_2017,Gagne2018_banyansigma,kounkel_covey_2019}. Furthermore, existing groups can be re-analyzed for previously missed members with these new kinematic data. One such example is the Tucana-Horologium Association (Tuc-Hor). Located within the solar neighborhood ($<$ 100pc), the core members of Tuc-Hor were first identified in the works of \citet{Torres_2000_first} and \citet{zuck_webb_2000_first}, and combined into one association by \citet{zuck_2001_combine}. The association is spread across a large area of the sky, making member identification more difficult than for compact groups like the Pleiades.

Fortunately, identifying young stars is possible in other ways. Stellar chromospheres are more active at younger ages as magnetic activity is driven by a vigorous dynamo \citep{dynamo_rotation_reiners_2012}. Observational signatures of youth can come from spectroscopic observations of excess \ha{} emission and high lithium abundance (Li), as well as X-ray or near ultraviolet (NUV) luminosity, which are associated with chromospheric activity \citep[e.g.,][]{xray_young_kin_malo_2014,nearby_young_galex_2013_rodriguez}, and have intrinsic spread across spectral types. \ha{} emission is linked to activity in M dwarfs and serves as a powerful age predictor, albeit less useful in F,G and early K stars where it is an absorption feature \citep[e.g.][]{kiman_2021_ha}. X-ray luminosity has been shown to be an excellent indicator of youth in young moving groups \citep{xray_young_kin_malo_2014}. Similarly, NUV has been shown to be a useful tool for distinguishing young stars from the field population \citep{nearby_young_galex_2013_rodriguez,Gagne2018_banyansigma,gagne_volans_2018}. 

The presence of a Lithium absorption line at 6707.8\,\AA\ is regarded as another age dating technique. Li is depleted in the lower convective boundary of stars during their lifetimes, but the rate at which it decreases is dependent on mass \citep{dantona_1994_liboundary}. The lithium depletion boundary, beyond which all Li has been depleted, has been used as an age-dating technique for young clusters and associations \citep[e.g.][]{dobbie_ic2602_Li_2010}. For stars warmer than the lithium depletion boundary, the lithium depletion rate on the main sequence appears to be correlated with rotation, as seen among K dwarf members of the $\sim$100~Myr Pleaides cluster \citep{bouvier_2018_li_plei} and Pisces--Eridanus stream \citep{ArancibiaSilva2020}. Therefore observations of the Li feature at 6708~\AA\ are an important metric for describing the overall population of an association. 

A number of techniques have been used to place constraints on the age of Tuc-Hor. \citet{kraus_2014_tuc_ha_li} considered the Li depletion boundary to determine an age of 40 Myr. This value was found to be consistent with an age estimate based on Bayesian analysis of member kinematics \citep{Crundall2019_age} as well as an isochronal age \citep{Bell2015_age}.

However, none of the previous age estimates for Tuc-Hor have incorporated gyrochronology \citep{barnes_2003}. Gyrochronology is another powerful independent age-dating technique \citep[e.g.][]{curtis_pis-eri,angus_2015,angus_2018_GP,timo_2015}. Older stars spin more slowly \citep{skumanich_1972}, which is thought to be caused by stellar winds interacting with stellar magnetic fields, reducing the star's angular momentum over time. These ``Skumanich-like'' relations for the spin down have been successful in estimating ages for solar-like stars older than hundreds of millions of years into field age, and has been revolutionized with space-based, long time baseline, high precision photometric missions such as Kepler \citep{kepler} and K2 \citep{k2}. These missions have enabled rotation period observations of field populations \citep{mcq2014} and coeval populations at a few hundred million to billions of years of age, to serve as benchmark groups to better understand the angular momentum evolution of stars \citep[e.g.][]{douglas2017,Rebull2016_plei,agueros_ngc752}.

More recently, the NASA Transiting Exoplanet Survey Satellite \citep[TESS;][]{TESS} has enabled rotation period measurements for structures that are dispersed over many degrees in the night sky. Associations that are spread over a large angular area make targeted observations challenging and expensive, as multiple fields are needed to cover all the targets. TESS is a powerful tool as it provides months of high precision time-series photometry for nearly the entire sky, and regularly returns to fields every other year-long Cycle. \citet{curtis_pis-eri} used TESS observations to measure the rotation periods for objects in the Pisces-Eridanus stellar stream and constrain its age more accurately when compared to the first isochronal study \citep{meingast_2019_pscieri}. For a moving group like Tuc-Hor, TESS provides the ideal data set to collect rotation periods from F stars down into the M dwarf regime.

Furthermore, recent studies of rapidly rotating young M's have uncovered an interesting new phenomenon within particular light curves. Originally called ``scallop-shaped objects'' in \cite{Stauffer2017_scallop1} due to the structure of the light curve being reminiscent of a scallop shell in profile, the term ``complex rotator'' is now more commonly used to describe these objects \citep{Gunther2020_scallop_spec}. They are characterized by sudden changes in brightness on short sub-hour timescales, for which there is no agreed upon explanation \citep{Gunther2020_scallop_spec}. These sudden changes repeat over cycles of revolution and are known to be stable for months at a time. TESS provides an opportunity to observe the evolution of these light curves between Cycle 1 and Cycle 3, years apart.

In this work we present the rotation rate distribution of \rotobs{} objects from Tuc-Hor. 
In Section \ref{memb_data} we discuss our membership list for Tuc-Hor, and we present the available TESS data, as well as the literature sources of youth indicators. In Section \ref{light_curve} we describe our process for identifying rotation periods in light curves derived from TESS full frame images. In Section \ref{sec:recovery} we discuss our efforts to account for contamination and present a recovery rate for rotation periods in our sample. In Section \ref{diag_1} we highlight interesting objects, including complex rotators in Section \ref{complex} and candidate binaries in Section \ref{binaries}. We describe our rotation period distribution for Tuc-Hor in Section \ref{sec:gyro} in context with Upper Scorpius and the Pleiades. In Section \ref{sec:diag_2} we combine our rotation periods with other indicators of youth to provide more context for Tuc-Hor's age. We end with a discussion in Section \ref{discuss} and conclude in Section \ref{conclusions}.

\section{Target Lists and Available Data}\label{memb_data}

The last census specific to Tuc-Hor, \citet{kraus_2014_tuc_ha_li}, generated a membership list for the group based on kinematic motion and photometric SED fitting. However, the authors note that systematic errors in the SEDs of early M dwarfs may have caused some members to be missed. Subsequently, Gaia EDR3 \citet{gaia_edr3} has delivered more accurate and precise parallaxes and proper motions for objects across the moving group, including into this faintest regime, and updated membership lists have been produced \citep[e.g.][]{banyan_2018_firstlook}. We are therefore using updated lists from the BANYAN $\Sigma$ Bayesian algorithm \citep[Gagn\'e priv. comm.; for algorithm details see][]{Gagne2018_banyansigma}. The updated lists are generated from membership probabilities drawn from Gaia EDR3 kinematics, but include a more thorough literature search for radial velocity measurements.

\subsection{Membership lists and catalog of signatures of youth}\label{memb}

After membership probabilities are generated from BANYAN $\Sigma$, objects are grouped into broad categories, including bona fide (BF), high-likelihood (HM), candidate members (CM), low-likelihood (LM), and rejected members. 
These classifications reflect the degree of certainty of objects being part of a known kinematic association. BF objects are often founding members of the association that have Galactic Cartesian position and velocity ($XYZ/UVW$) values that define its extent, and also show a clear sign of youth (e.g., X-ray emission, color-magnitude diagram position). HM stars have high membership probabilities ($>$90\%), due to a close match to the kinematic values of the group and also have signs of youth. An object might also be labeled an HM instead of BF if they are missing either a kinematic measurement (e.g., RV) or a clear sign of youth, even if all other values are a good match to the association's values. CM objects are those that are missing additional measurements (e.g. RV and youth indicator), or whose probability of membership is $<$90\%. LM are objects which have some probability of membership but which is mostly considered negligible, and rejected members are those that were once thought to be part of the group but subsequently were ruled out due to updated kinematic measurements.

In this work, we used the BF, HM, and CM lists to create our Tuc-Hor sample and ignored the low-likelihood and rejected members. This led to an initial list of \fulltuchor{} objects. Including CMs is important as they extend to redder objects than the BF and HM lists. These intrinsically fainter objects may have a higher chance of being contaminants or interlopers, but by including these less certain objects we can probe down into the lowest mass stars, and even brown dwarfs. Once we combined the objects into a single list we disregarded their prior classification and considered them all equally. 

For our final Tuc-Hor sample we queried a database that includes available literature measurements for various signatures of youth (Gagn\'e priv. comm.). The database draws from the following:
\begin{itemize}
    \item \ha{} equivalent width measurements are drawn from \citet{schneider_2019_ha, kraus_2014_tuc_ha_li,torres_2006,shkolnik_2017_ha,white_2007}. 
    \item Li equivalent width measurements come from \citet{shkolnik_2017_ha,torres_2006,white_2007,kraus_2014_tuc_ha_li}.
    \item NUV photometry comes from a cross-match with GALEX \citep{GALEX_2005}. 
    \item X-ray photometry comes from a cross-match with ROSAT \citep{ROSAT_2016}. 
\end{itemize}
We list the Gaia~EDR3 R.A., Dec, $G$, $G_{\rm BP}$, $G_{\rm RP}$ for the objects in Table~\ref{table:sources}, along with their BANYAN $\Sigma$ classification, and the available measurements for signatures of youth. We did not correct any of the photometry for reddening, as we assumed its effect on the dispersed moving group was negligible.

\subsection{TESS Observations}\label{data}

NASA's Transiting Exoplanet Survey Satellite \citep[TESS;][]{TESS} is an all-sky mission that images sectors of the sky for $\sim$28 days each, roughly covering a hemisphere in a 13 sector-long cycle. Even a single sector baseline is sufficient for measuring stellar rotation for Tuc-Hor objects, as the rotation period for 40 Myr objects should not exceed 10 days.(e.g. in the 120 Myr Pleiades 98\% of the objects rotate more rapidly than 10 days) \citep[e.g.][]{Rebull2016_plei}. 

\begin{figure}
\begin{center}
\includegraphics[width=1.0\linewidth]{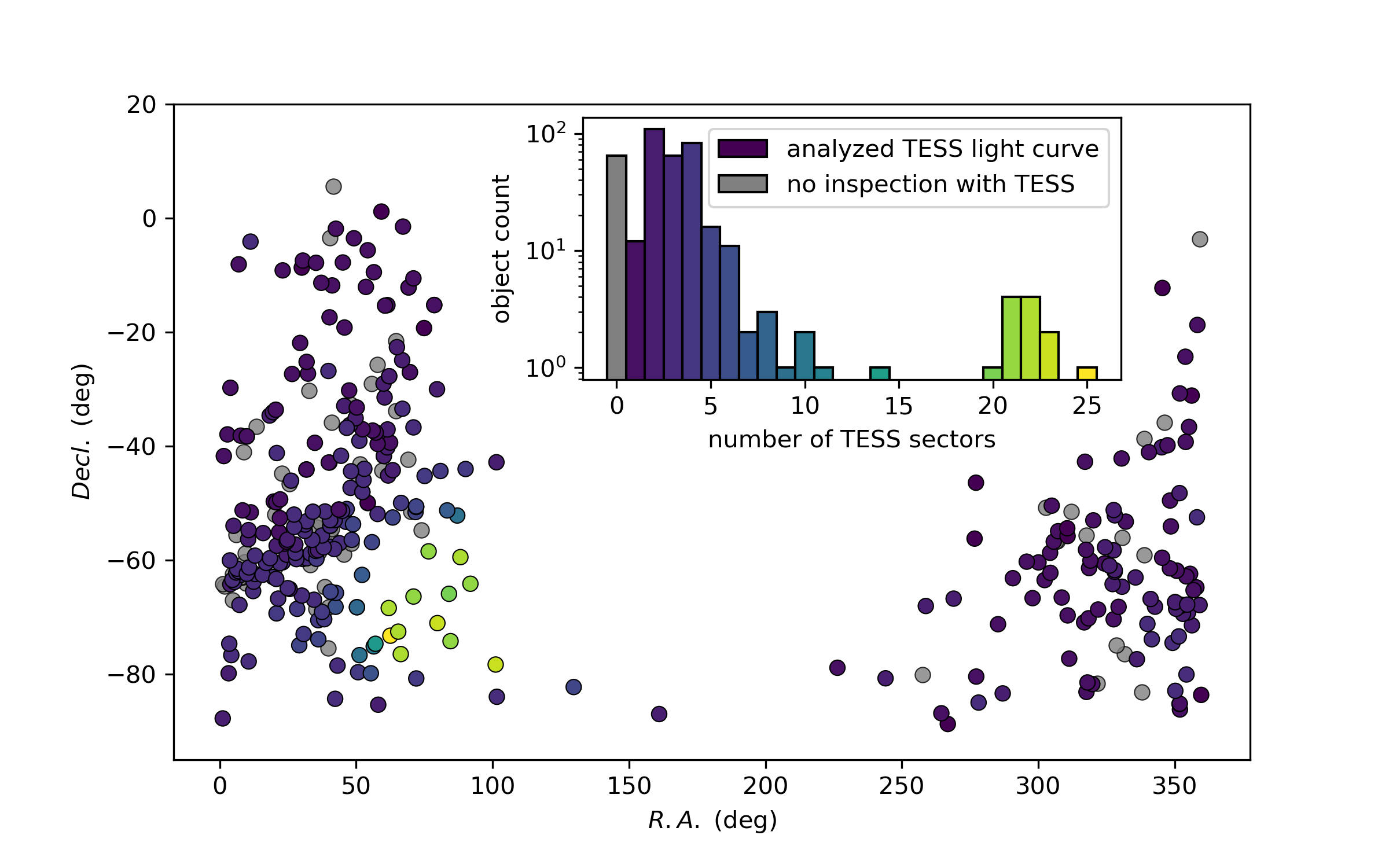}
\caption{The sky position of the objects in our Tuc-Hor sample. Points are color coded by the number of sectors available in TESS, or grey if they were not analyzed with TESS. Inset is a histogram with the count of the number of sectors we analyzed for each object in our sample. Of the 69 targets that we did not inspect, 11 were not observed by TESS, 7 were too blue to be F stars, and the other 51 were fainter than our $G$ limit of 18 or missing Gaia photometry.}
\label{fig:sector_hist}
\end{center}
\end{figure}

We began by querying MAST for Full Frame Image (FFI) cutouts based on Gaia DR2 right ascension and declination values listed in Table~\ref{table:sources}. We downloaded 40$\times$40 pixel cutouts from the FFI for each sector available for our targets. Additionally, we set an upper Gaia \gbr\ color limit $<$ 0.5 to filter out members with spectral types earlier than F stars, as well as a brightness limit of Gaia $G < 18$~mag. This reduced our full sample of \fulltuchor{} to \withtess{} objects with TESS observations. \footnote{We originally set our magnitude limit to $G < 16$, based on our experience working on older clusters surveyed by TESS with ages of $\sim$200--400~Myr. However, we found in this study that we could confidently detect rapid rotation periods for much fainter stars, probably because the light curves capture many more complete revolutions and the photometric amplitudes are much larger at this age and rotation rate. For this reason we expanded our faintness limit to $G < 18$~mag.}
The number of TESS sectors available for each object are listed in Table ~\ref{table:sources}, and are presented as a histogram in Figure~\ref{fig:sector_hist}.

On average, four TESS sectors are available up to June 2022 for our targets. 
Due to our magnitude cuts and the TESS survey footprint, 69 targets on our list were not inspected for rotation: 11 were not observed yet by TESS, 7 were too blue to be F stars, and 55 were fainter than $G$ of 18 $mag$, which was our faintness limit or missing photometry.

\begin{deluxetable*}{hccccccchhhhhhchhchhchchc}
\tabletypesize{\scriptsize}

\tablecaption{Tuc-Hor Sample \label{table:sources}}
\tablewidth{0pt}
\tablehead{
\nocolhead{Gaia DR2 Source ID}& \colhead{Gaia DR3 Source ID}& \colhead{Gaia R.A.} & \colhead{Gaia Decl.}&  \colhead{BANYAN $\Sigma$ class}& \colhead{$G$} & \colhead{$G_{\rm BP}$} & \colhead{$G_{\rm RP}$} & \nocolhead{$G$ flux over error} & \nocolhead{$G_{\rm BP}$ flux over error} & \nocolhead{$G_{\rm RP}$ flux over error}&  \nocolhead{Parallax}& \nocolhead{Parallax over error}& \nocolhead{ruwe}& \colhead{EW \ha{}} & \nocolhead{EW \ha{} Error}&  \nocolhead{EW \ha{} Source} &  \colhead{EW Li} & \nocolhead{EW Li Error}&  \nocolhead{EW Li Source} &  \colhead{X-Ray Flux} & \nocolhead{X-Ray Error} &  \colhead{NUV Mag} & \nocolhead{NUV Error} & \colhead{Num TESS Sectors}\\
 & & (deg)& (deg)& &(mag)& (mag)& (mag)& & & & (mas)& & &(\AA) & (\AA)&  & (m\AA)& (m\AA) &  & (flux) & (flux) & (mag) & (mag) &  }
\startdata
6478258087046528 &    6478258087046528 &  41.561207 &   5.592393 &       BF &  7.754 &   8.008 &   7.333 &             1949.25670 &              549.040830 &               553.39340 & 17.530643 &           734.948400 &  0.965007 &      $\cdots$ &          $\cdots$ &                                                $\cdots$ &       $\cdots$ &           $\cdots$ &                                                $\cdots$ &        0.0 &       0.0 &      $\cdots$ &      $\cdots$ &          0 \\
2308129860256072448 & 2308129860256072448 &   2.814451 & -37.949093 &       CM & 15.894 &  18.436 &  14.465 &              765.50810 &               24.785864 &               511.57843 & 22.224268 &           372.931550 &  1.082642 &      $\cdots$ &          $\cdots$ &                                                $\cdots$ &       $\cdots$ &           $\cdots$ &                                                $\cdots$ &        $\cdots$ &       $\cdots$ &      $\cdots$ &      $\cdots$ &          2 \\
2311548448064869120 & 2311548448066274944 & 355.288152 & -36.638621 &       CM & 14.375 &  16.343 &  13.033 &             1080.83830 &              145.788390 &               617.04395 & 25.783608 &           414.547820 &  0.943507 &      $\cdots$ &          $\cdots$ &                                                $\cdots$ &       $\cdots$ &           $\cdots$ &                                                $\cdots$ &        $\cdots$ &       $\cdots$ &   22.561 &    0.317 &          2 \\
2320267025518037760 & 2320267025518037760 &   3.903525 & -29.767171 &       BF & 12.833 &  14.539 &  11.572 &             1144.73130 &              153.652330 &               639.48400 & 27.578012 &           776.719000 &  1.225291 &    -6.32 &         0.05 &        schn19Schneider et al. (2019) ACRONYM. III. &       $\cdots$ &           $\cdots$ &                                                $\cdots$ &        $\cdots$ &       $\cdots$ &   20.361 &    0.131 &          2 \\
2326515034702293760 & 2326515034702293760 & 356.256315 & -31.135017 &       CM & 14.754 &  16.500 &  13.496 &             1472.63110 &              171.241640 &              1264.57920 & 25.486082 &           819.732200 &  1.084068 &      $\cdots$ &          $\cdots$ &                                                $\cdots$ &       $\cdots$ &           $\cdots$ &                                                $\cdots$ &        $\cdots$ &       $\cdots$ &   23.084 &    0.469 &          1 \\
2328825933265980544 & 2328825933265980544 & 351.934532 & -30.761420 &       CM & 13.641 &  15.431 &  12.365 &              770.62665 &              141.893230 &               314.69556 & 22.559446 &           941.330140 &  1.209996 &      $\cdots$ &          $\cdots$ &                                                $\cdots$ &       $\cdots$ &           $\cdots$ &                                                $\cdots$ &        $\cdots$ &       $\cdots$ &   21.622 &    0.204 &          2 \\
2380973085416335104 & 2380973085416335104 & 353.953615 & -24.319779 &       CM & 12.470 &  13.931 &  11.284 &             1115.60470 &              274.719760 &               646.79944 & 29.774436 &          1037.256100 &  1.389660 &      $\cdots$ &          $\cdots$ &                                                $\cdots$ &       $\cdots$ &           $\cdots$ &                                                $\cdots$ &        $\cdots$ &       $\cdots$ &   21.900 &    0.533 &          2 \\
2390232107194169088 & 2390232107194169088 & 358.356979 & -18.744924 &       CM & 14.768 &  16.996 &  13.400 &             1369.41830 &              130.597980 &               683.76890 & 27.375649 &           726.427200 &  1.054297 &      $\cdots$ &          $\cdots$ &                                                $\cdots$ &       $\cdots$ &           $\cdots$ &                                                $\cdots$ &        $\cdots$ &       $\cdots$ &      $\cdots$ &      $\cdots$ &          2 \\
2430275225461072128 & 2430275225461072128 &   6.939477 &  -8.101572 &       CM & 15.290 &  17.773 &  13.872 &             1210.61390 &               97.410150 &              1019.56885 & 24.372852 &           340.229800 &  1.427102 &    -5.53 &         0.05 &                                             shko17 &     650.0 &          50.0 &                                             shko17 &        $\cdots$ &       $\cdots$ &   22.087 &    0.421 &          2 \\
2448022786243469696 & 2448022786243469696 & 359.267682 &  -3.632460 &       HM & 14.009 &  15.567 &  12.690 &              254.08305 &              426.372960 &              1288.35050 & 18.948042 &            26.781748 & 32.729490 &    -4.60 &         0.10 & krau14Kraus et al. (2014) Spectroscopy of Tuc-H... &      24.8 &          50.0 & krau14Kraus et al. (2014) Spectroscopy of Tuc-H... &        0.0 &       0.0 &   20.438 &    0.220 &          0 \\
\enddata
\tablecomments{Objects in our Tuc-Hor membership list. We include Gaia DR 2 information as well as available measurements for various indicators of youth. Sources for these measurements are discussed in Section ~\ref{memb}. We also include the number of TESS sectors for which the object was observed based on it's R.A. and decl. This is an example of the \fulltuchor{} row table in the full online version.}
\end{deluxetable*}

\section{Light Curve Analysis}\label{light_curve}

This work utilizes an open-source Python GUI that is under development\footnote{The associated code can be found at \url{https://github.com/SPOT-FFI/tess_check}} to manage object lists, download and create light curves, and measure rotation periods.

\begin{figure}
\begin{center}
\includegraphics[width=1.0\linewidth]{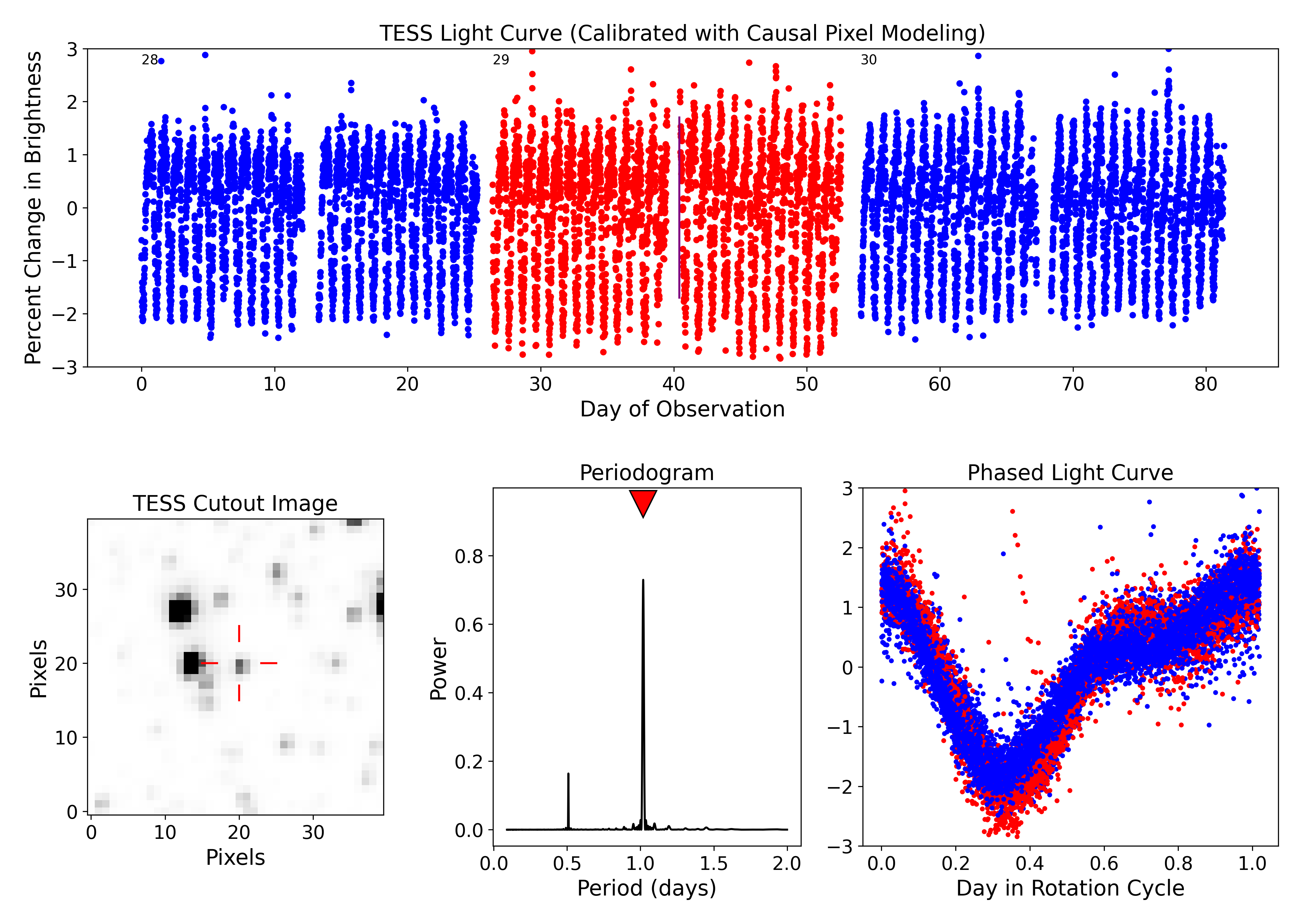}
\caption{The four figure panels for Gaia DR3 4668269708587377664. The top figure is the full light curve for the object, with sectors plotted in alternating red and blue. The purple vertical line in the middle denotes the amplitude across the light curve. From left to right along the bottom are the image of the TESS FFI cutout, a Lomb-Scargle periodogram of the light curve, and the light curve phase folded on the period with highest power from the periodogram. The description of how we used each figure for inspection is in the text. A figure set of panels for all the objects in our Tuc-Hor sample is available in the online journal.}
\label{f:tesscheck}
\end{center}
\end{figure}

From the 40$\times$40 pixel FFI cutouts, we generated two light curves for each sector; one using the causal pixel modeling (CPM) implemented in the Python \texttt{unpopular} package \citep{hattori_cpm_2021} and another using simple aperture photometry (SAP) with a two-pixel aperture centered on the object \citep[here, the assumption is that TESS systematics are not as significant as the large photometric modulation amplitudes observed in young stars;][]{curtis_pis-eri}. After experimenting on a variety of light curves with a range of magnitudes we found that CPM light curves were preferable for objects with $G >$ 10~mag, and the SAP light curves were preferable for those with $G \leq$ 10~mag, but we cross-checked our choice in border cases if needed. For example, some sectors would show issues with the detrending process, most likely due to either poor pixels in the case of the CPM method, or TESS systematics affecting the SAP method. We then visually inspected all available sectors and subsequent light curves for each object. During our visual inspection we referenced four figure panels for each object; for example, see Figure~\ref{f:tesscheck} (panels for all targets are available as a figure set in the online journal).

Each figure includes the full light curve used in the analysis, the Lomb-Scargle periodogram for that light curve, a phase folded light curve using the period with highest power, and the FFI cutout centered on that object. During the process of inspection we occasionally adjusted the range of the percent change in brightness to avoid flares, or limited the range of period space searched by the periodogram when the maximum peak was clearly at a short period. We also chose to reject particular sectors for some objects. If a sector showed an issue caused by the detrending method, it would be removed during the inspection and not included in the period calculation. If a rotational variability signal was seen in even a single sector, we approved the period. Most objects were variable in every sector available.

\begin{deluxetable*}{hrrllrlr}
\tablecaption{Tuc-Hor Rotation Periods\label{table:rotation}}

\tablehead{
\nocolhead{DR2 Name} & \nocolhead{DR3 Name} & \colhead{Period} & \colhead{Class} & \colhead{Notes} & \colhead{Amplitude} & \colhead{Sectors Used} & \colhead{Flares Flag}\\
& & (days) &  &  & (ppm) &  }

\startdata
6478258087046528 &    6478258087046528 &   -1.000000 &     $\cdots$ &                                No TESS Data &      $\cdots$ &          $\cdots$ &    NA \\
2308129860256072448 & 2308129860256072448 &    0.100350 &    Good in both sectors, more obvious in 2, maybe no flare? & 0.009412 &       ('2',) &       1 \\
2311548448064869120 & 2311548448066274944 &    0.881737 & Publish &                           huge flares in 29 & 0.011349 &       ('2',) &       1 \\
2320267025518037760 & 2320267025518037760 &    0.836228 & Publish &                                         $\cdots$ & 0.027612 &  ('2', '29') &       1 \\
2326515034702293760 & 2326515034702293760 &    0.291015 &    Good &                                   big flare & 0.014139 &      ('29',) &       1 \\
2328825933265980544 & 2328825933265980544 &    0.537952 & Publish &                    other period is harmonic & 0.016045 &      ('29',) &       1 \\
2380973085416335104 & 2380973085416335104 &   99.000000 &    Flat &                                       & 0.005258 &      ('29',) &       0 \\
2390232107194169088 & 2390232107194169088 &    0.615460 & Publish &                              no flares in 2 & 0.015511 &      ('29',) &       1 \\
2430275225461072128 & 2430275225461072128 &    0.476296 &    Good &                     faint but there in both & 0.011575 &      ('30',) &       0 \\
2448022786243469696 & 2448022786243469696 &   -1.000000 &     $\cdots$ &                                No TESS Data &      $\cdots$ &         $\cdots$ &    NA \\
\enddata
\tablecomments{The rotation periods measured for the Tuc-Hor sample. We list the period measured, our decision based on visual inspection to whether the period is acceptable, and the amplitude of variability, the TESS sectors used to determine the period, and a binary flag for if flares were observed in any part of the light curve. A full description of the Class labels can be found in the body of the text.}
\end{deluxetable*}

Table ~\ref{table:rotation} reports our rotation periods. We list which sectors were used to determine the rotation period for each object. Objects lacking TESS data are still included and their periods are reported as $-1$ and a note is made in the ``Notes'' column. Objects with an exceptionally strong period were classified as ``Publish'', and represent objects with light curves with distinct and obvious star spot induced rotation periods. ``Good'' are objects with recovered rotation periods but are not as strikingly obvious as ``Publish'' objects. We include both ``Publish'' and ``Good'' as successfully recovered rotation periods. While the difference is somewhat subjective, we include it as it may be useful for future light curve analysis work. This table also includes objects where no rotation period was evident. They are listed in the class column as either ``Flat'', meaning their light curve showed no rotation period but otherwise looked astrophysical, or ``Ignore'', meaning their light curve looked to be dominated by non-physical systematics or detrending errors. We discuss the frequency of these non-detections in Section ~\ref{sec:recovery_rate}.

%

\section{Rotation period recovery}\label{sec:recovery}

\subsection{Contamination in TESS}\label{contamination}

TESS has 21$\times$21$''$ pixels, which are rather large compared to Kepler's 3.98'' pixels or detectors commonly used for ground-based observations ($\sim$ 0.1$"$). Given this large pixel size, we wanted to ensure that our recovered (or non-recovered) periods were free of contamination from the light of nearby sources. We expect young stars to vary significantly in brightness due to their high levels of magnetic activity \citep[e.g.,][]{Rebull2018_usco, Rebull2016_plei}. Therefore we assume a lack of such a signal is either due to contamination by a nearby source or due to a physical attribute of the star (e.g.,m pole-on orientation, in a quiescent activity state, an older age interloper to the moving group).
To define contamination by a nearby source we considered the Gaia magnitude ($G$) of the source, the magnitude difference with the nearby object ($\Delta G$), and the separation between target and closest neighbor ($\Delta$ $\alpha$) in arcsec. To describe the influence of these factors, we started by examining each target from the full Tuc-Hor sample for other Tuc-Hor sources that could be a contaminate. We began by searching around each Tuc-Hor member in a ten arcminute radius and matching proper motions to find the widest swath of nearby Tuc-Hor members. This Tuc-Hor--Tuc-Hor contamination sample is particularly powerful since we have good reason to suspect that both objects should have a recoverable period given their age.

\begin{deluxetable*}{rrrrrrll}
\tablecaption{Tuc-Hor--Tuc-Hor Contamination.\label{THA-THA}}

\tablehead{
\colhead{Obj 1 DR2 Name} &  \colhead{Obj 2 DR2 Name} & \colhead{Obj 1 $G$ mag} & \colhead{Obj 2 $G$ mag} & \colhead{$\Delta G$} & \colhead{$\Delta$ $\alpha$} & \colhead{Period 1} & \colhead{Period 1}\\
& & (mag) & (mag) & (mag) & (arc sec) & (days) & (days)}

\startdata
4832672020067562496 & 4832672020067562624 &   13.189 &    7.474 &           5.715 &       4.208 &      -- &   2.538 \\
5000558409016727296 & 5000558443376465408 &   12.828 &   11.362 &           1.466 &      20.310 & (6.214) & (1.049) \\
4714764481913306496 & 4714764447553568640 &    7.370 &    9.857 &           2.487 &      94.583 &   1.718 &   4.026 \\
4742040410461492096 & 4742040513540707072 &    9.736 &   11.076 &           1.340 &      22.876 & (0.519) & (0.559) \\
4842275841819363200 & 4842275837523665664 &    8.221 &   12.287 &           4.066 &       9.180 &   3.487 &      -- 
\enddata
\tablecomments{Pairs with potential contamination within our Tuc-Hor sample. We list the Gaia DR2 names and magnitudes for each object, as well as the magnitude difference and separation for the objects. Available light curves for each pair were visually inspected in an attempt to assign periods to the objects. If after visual inspection we could confidently assign a period, we list it in the appropriate Period column. If we are uncertain about one period we place it in parentheses, and if we are uncertain for which object either period should be assigned we randomly assign the period and list both in parentheses. This is the first five of twenty such pairs.}
\end{deluxetable*}

In all, we found that there were 20 pairs and one triple with separations $<$ 10$'$ within the full Tuc-Hor membership list. We identify them in Table~\ref{THA-THA} with their magnitudes, separations, and period assigned after visual analysis (described below). The separation for these pairs range from within $\sim$3$''$, to $\sim$360$''$. 
For the contamination analysis we discarded the triple because we could not find a rotation period due to two of the stars being bright A stars ($G$ = 4.27 and 5.33~mag) compared to the third component M dwarf ($G$ = 12.25~mag; Gaia DR2 Source IDs are 4856719713756946176, 4856719713756945664, and 4856719640741737216, respectively). 

The final light curves we use for analysis (both CPM and SAP, see Section~\ref{light_curve}) begin from establishing a single target pixel that contains the object's RA and Dec. The TESS pointing is defined over large areas of sky, thus the position of a star on the grid of pixels will vary from sector to sector. This meant that for pairs with objects that were under a certain separation threshold ($\sim 21''$) the light curves for both stars were generated from the same target pixel for each sector and were identical, while in other pairs with separation $\gtrsim 21''$ we had multiple light curves to analyze.

We sorted the 20 pairs into 5 categories: \begin{enumerate}

    \item pairs where we had two periods and we could assign a period to each member with confidence.
    \item pairs where we had two periods but it was unclear to which star they should be assigned.
    \item pairs with one obvious period, and a potential secondary period
    \item pairs with only one period which we could assign to a specific member with confidence.
    \item pairs with only one period and it was unclear to which star it should be assigned.
\end{enumerate}

\begin{figure}
\begin{center}
\includegraphics[width=1.0\linewidth]{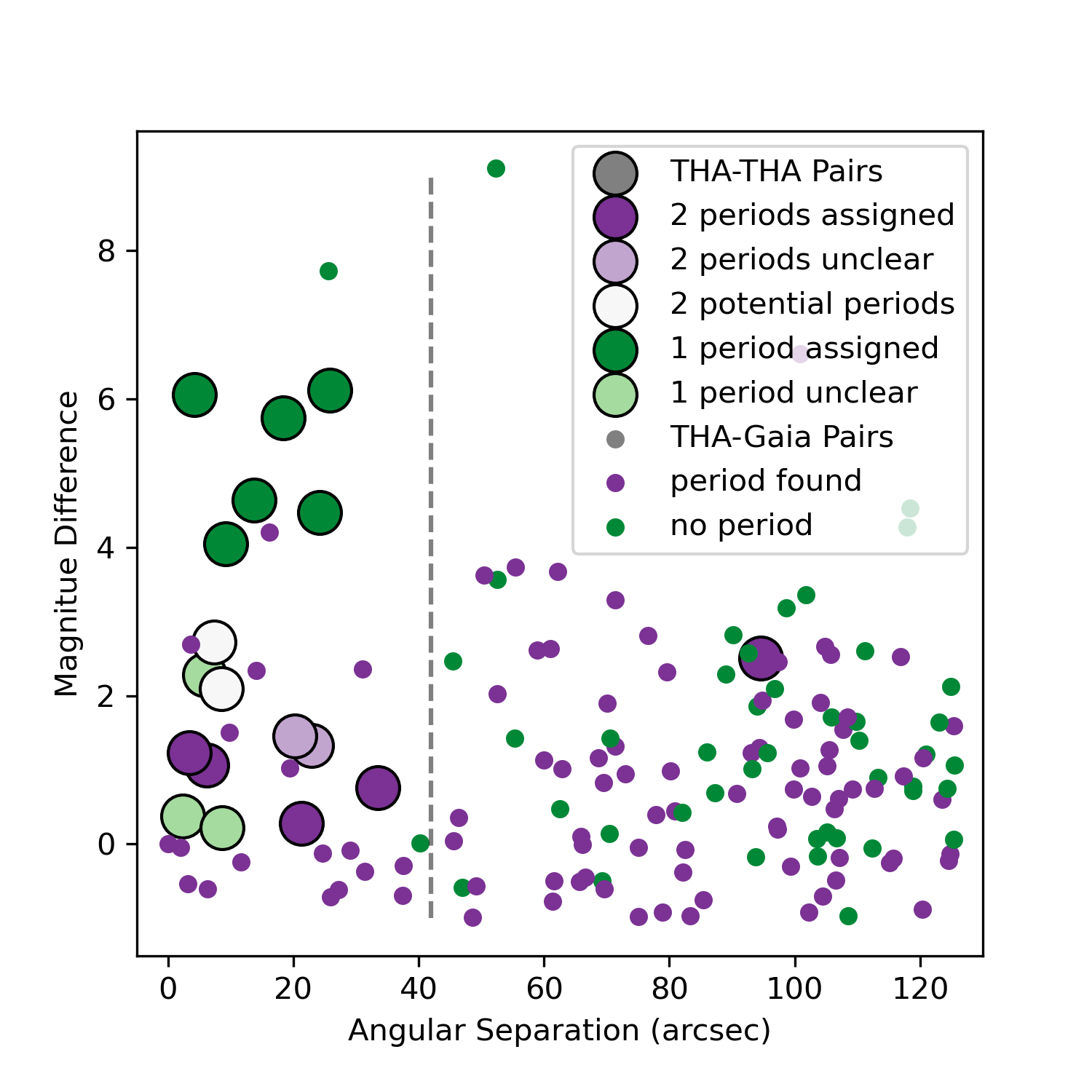}
\caption{Angular separation versus magnitude difference for Tuc-Hor--Tuc-Hor pairs (large circles) and Tuc-Hor-Gaia pairs (small dots, ). THA is used as Tuc-Hor for the purpose of the legend. Tuc-Hor--Tuc-Hor Pairs are color coded based on the status of rotation periods after visually inspecting available light curves for each pair, and presented in the order of the catgeories described in Section ~\ref{contamination}. Purple and white generally denotes that we found the expected number of periods (two for the case of Tuc-Hor--Tuc-Hor pairs, one for Tuc-Hor--Gaia), and green that a rotation period is missing (only one for the case of Tuc-Hor--Tuc-Hor pairs. The vertical dashed line represents 42$''$ (two TESS pixels) of separation. For Tuc-Hor--Tuc-Hor pairs we find that within 2 TESS pixels of separation and within a 2 magnitude difference we are often able to recover the expected number of periods.}
\label{fig:contamin}
\end{center}
\end{figure}

We were able to confidently assign  periods to both components (Category 1) when the sources were separated by at least 1 TESS pixel and no more than 2 mags. The one exception to this was Gaia DR2 6387058411482257536 (DS Tuc A), a known planet host. The planet was discovered using TESS light curves \citep{thyme_dustuc_2019}, but there was a previous literature rotation value \citep{kiraga_2012_dustucrot} and so we could assign the other period to Gaia DR2 6387058411482257280 (DS Tuc B) even though it was only a few pixels away. We also note that Category 5 pairs --- where there was only one period detected and it was unclear which source it was coming from ---  contained some of the faintest objects in our sample, and likely pushed against our recovery limit (see Section ~\ref{sec:recovery}).

Based on the results we found from analyzing the Tuc-Hor--Tuc-Hor pairs we made the following rules for considering contamination with background sources. First, multiple periods could be found as long as objects were separated by multiple TESS pixels, dependent on the magnitudes of the objects. Sources with Gaia $G$ $<$ 10 had an increasingly larger pixel footprint, their light spreading over 3 or more pixels. However, approximately 90\% of the objects within the sample are G $>$ 10 mag. Next, a neighbor becomes likely to wash out a signal when it is approximately two magnitudes brighter. All the objects for which we were able to recover two periods had a magnitude difference $<$2~mag or had a separation $\geq$ 42$''$ (about two TESS pixels).

Using this understanding we next searched for any Gaia source within a six TESS pixel radius (126$''$) radius of a Tuc-Hor object. We plot their $\Delta$ mag and $\Delta$ $\alpha$ alongside that of the Tuc-Hor--Tuc-Hor pairs in Figure ~\ref{fig:contamin}. The small purple dots represent objects with a recovered a period and the green dots those without a recovered period. Based on our assumption that a generic Gaia star is likely a few billion years old and likely quiescent, we attribute all of the detected periods to be from the Tuc-Hor objects. For angular separations $>$ 42$''$ objects with periods found are interspersed with objects with no period found. We therefore conclude that there is unlikely contamination, and that objects without a period truly do not have one detectable with TESS. We found a total of six of our objects that had a nearby source two magnitudes brighter and within two TESS pixels. From visual inspection of these light curves we conclude that it is likely that the potential signal was diluted by the brighter contaminant. We do not consider them in terms of recovery rates and effectively remove them from the sample.

We end this discussion by commenting that despite the large TESS pixels, our analysis shows that only a small number of sources had to be removed from consideration due to contamination. There are likely six Tuc-Hor objects washed out by another Tuc-Hor object, and only six that we remove due to contamination from a non member. This means that after light curve inspection and considering contamination, we have \rotobs{} objects with rotation period measurements.


\subsection{Recovery Rate}\label{sec:recovery_rate}

To calculate our rate of recovery for rotation periods across the Tuc-Hor sample we exclude all of the potentially contaminated objects discussed in Section ~\ref{contamination}. To summarize, our contamination suggested that 12 objects from the \withtess{} objects with TESS observations were potentially being dominated by the light from nearby stars to the point that a rotation signal could be obscured. We intend this discussion then to be based on a recovery rate for Tuc-Hor objects that are free of contamination.

\begin{figure}
\begin{center}
\includegraphics[width=0.8\linewidth]{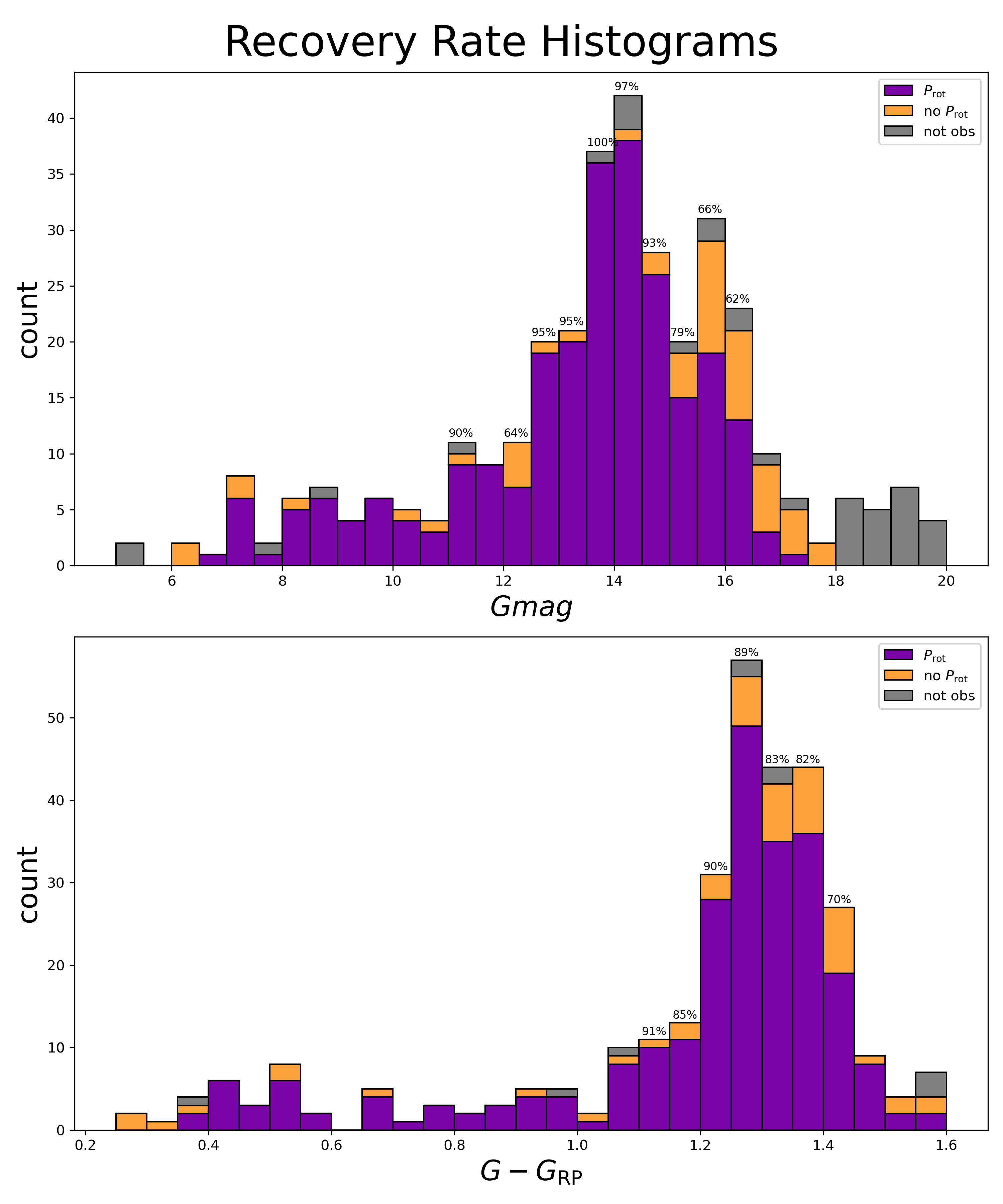}
\caption{\textit{Top} Histogram showing the distribution of target Tuc-Hor objects with recovered rotation rate across Gaia $G$ magnitude. We find that at 15.5-16 Gaia $G$ we begin to have a significant number of non-detection, suggesting that this is perhaps a point where rotation signals are competing with systematic noise. However we are still able to detect a period in object G mag $=$ 17.22. \textit{Bottom} Histogram showing the distribution of target Tuc-Hor objects with recovered rotation rate across \grp{}. When comparing to the top panel, we see that the non-detection is more evenly distributed across color, but with a larger percentage among \grp $\geq$ 1.2. For both top and bottom recovery percentages do not include objects not observed with TESS (grey).}
\label{fig:recover_hist}
\end{center}
\end{figure}

Figure~\ref{fig:recover_hist} presents histograms of the recovery rate for our full \fulltuchor{} sample for both G mag in the top panel, and \grp{} in the bottom panel. We include the objects not observed with TESS in grey for demographic purposes, while the recovery rate of \recovrate{} that we report is based on just those \withtess{} objects with TESS observations. In the top panel, each bin was 0.5 magnitudes in width, while the bottom has a bin spacing of 0.05 in \grp{}, and were chosen to conveniently divide up the range. It is important to note that of the 340 objects that appear in the top panel, only 309 appear in the bottom. This is due to certain objects not having good Gaia $G_{\rm RP}$ due to being too bright or too faint, and therefore not having a \grp{} value. The membership list is presented as either having a rotation period measured ($P_{\rm rot}$), having a light curve without a measurable rotation period ($no P_{\rm rot}$), or not having a light curve at all ($not obs$). We did not attempt to create light curves for objects fainter than G mag $=$ 18, so they are all considered $not obs$. Other $not obs$ are due to chance position of stars outside of the field of view of any TESS sectors.

For the top panel, the majority of the objects for which we do not observe a rotation rate are between 15.5 mag $\leq$ G $\leq$ 18 mag. We believe this increase in non detection is because we are operating at the faintness limit of TESS and only sensitive to the highest amplitudes for these objects. This is to be expected, but it is worth noting that it is exacerbated by the switch to 10 minute full frame cadence in Cycle 3. The shorter integration time means more systematic signal is introduced per data point, which can overpower the signal in the faintest objects. The faintest object for which we measure a rotation period was Gaia DR2 6362385542354403456 whose Gaia $G$ mag $=$ 17.22. 


In the bottom panel of Figure~\ref{fig:recover_hist}, the sample is presented in terms of \grp{} color and the non rotating objects ($no P_{\rm rot}$) are more evenly distributed across the full spectrum. As mentioned above, a significant number of the objects are missing from this plot, as they do not have a $G_{\rm RP}$ detection available in Gaia DR 2. We find that for even the reddest colors there are some objects that are recovered, while $no P_{\rm rot}$ quickly became a dominating percentage at fainter G Mag in the top panel. This is consistent with the expectation for detectable activity across color and spectral type for this age, and that $no P_{\rm rot}$ objects are generally due to a limiting magnitude. 

The overall high recovery rate of \recovrate{} is consistent for an association of this age. It is comparable to the fraction of recovered periodic members listed in \citet{Rebull2018_usco} for the Pleiades (92 \%) and Upper Scorpius (86 \%). We discuss the significance of the non-detection of a rotation on the membership probability of a given object in Section ~\ref{discuss}.

\section{Diagnosing the Sample}\label{diag_1}

\subsection{Complex Rotators}\label{complex}

\citet{Stauffer2017_scallop1} used K2 light curves to identify a class of rapidly rotating low mass stars in $\rho$ Ophiuchus and Upper Scorpius with complex light curve morphology that they dubbed ``scallop-shells''. The sample was expanded in \citet{Stauffer2017_scallop2} which included sources from the Taurus star forming region. Additional examples of these objects have been found in TESS, \citep[e.g.,][]{Zhan2019_scallops_tess,Stauffer2020_scallop_lcen_crux_ucen_lup}, and several sources have been followed up with ground based observatories in \citet{Gunther2020_scallop_spec}. A feature of these objects is their complex light curve structure that is maintained over the duration of observation. Consequently we choose to refer to these sources for the remainder of the text as ``complex rotators''. With the K2 discoveries, the length of observation was $\sim$80 days. For TESS discoveries, the average was $\sim$27 days but also up to one year for at least one object discovered in the Continuous Viewing Zone (CVZ) \citep{Gunther2020_scallop_spec}. The mechanism that causes the dramatic structure in the light curve has not been agreed upon. The changes in brightness can not be driven  by star spots alone as they are too sudden, occuring over the course of an hour or well within the rotation period of the object. Furthermore the observed stability of the light curve shape implies that the occulting structure/material be stable over the course of weeks to months. (See \citet{Gunther2020_scallop_spec} and Section \ref{complex_discuss} for more details) While starspots can exist on similar lifetimes, they usually have some evolution over that time frame, and are not capable of making such sharp features by themselves.

Of the ten targets focused on by \citet{Gunther2020_scallop_spec}, six of them are listed as Tuc-Hor members, based on \citet{Gagne2018_banyansigma} membership lists. In addition to the six known complex rotator objects, we identified three additional objects that have complex morphology in the 30 minute cadence light curves derived from TESS full frame images. \citet{Gunther2020_scallop_spec} showed that 30 minute cadence light curves will obscure details in the complex structure of these rapidly rotating objects. We therefore downloaded the two minute cadence PDCSAP files from MAST for all nine Tuc-Hor complex rotators. Each object has at least one light curve from both TESS Cycle 1 and Cycle 3, except for TIC 65347864 which only has one sector of two minute cadence in Cycle 1.

We comment on the light curve morphology over months and days when comparing successive sectors, as well as over years when comparing between Cycles in Section ~\ref{complex_comment}. We discuss the implications for the cause of this phenomenon in Section ~\ref{complex_discuss}

\subsubsection{Complex rotator light curves between TESS Cycle 1 and Cycle 3}\label{complex_comment}

\begin{figure}
\begin{center}
\includegraphics[width=0.4\linewidth]{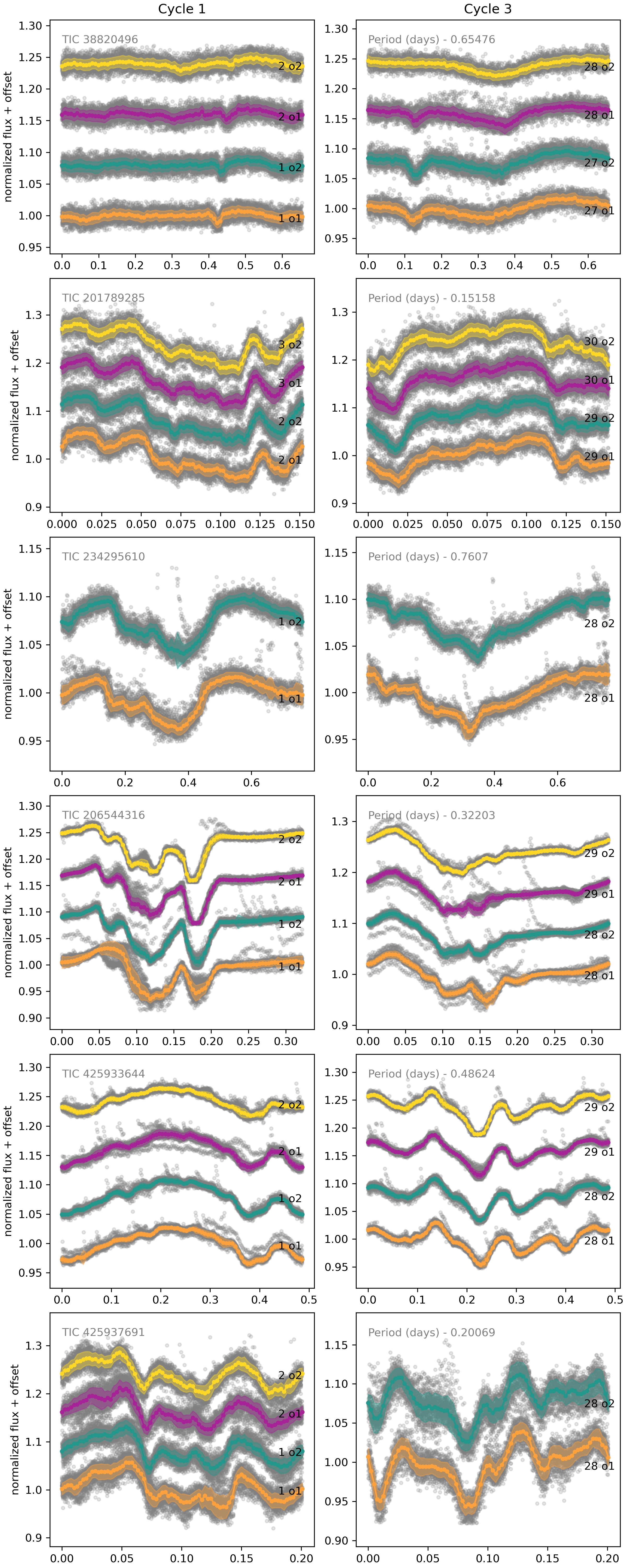}
\caption{Phase folded light curves from the two minute cadence TESS data for each sector of the previously identified complex rotators identified listed in \citet{Gunther2020_scallop_spec}. Both orbits in each sector are plotted separately, and the raw PDC SAP data are grey points. The first column are those from Cycle 1, and the second from Cycle 3. The phase position in Cycle 1 and Cycle 3 are not to be considered the same, as explained in the text. On top of each light curve is a median phase folded light curve with a shaded area of 1 standard deviation. We note a change in features for all objects from Cycle 1 to Cycle 3, which constrains the stability of this phenomenon to timescales less than a year.}
\label{fig:scallop_prev}
\end{center}
\end{figure}

We present phase folded median light curves for the six previously known complex rotators in Figure~\ref{fig:scallop_prev} using periods consistent with \citet{Zhan2019_scallops_tess}. We create a median light curve for the first and second half of each sector that the object appears, and present Cycle 1 sectors in a separate column from Cycle 3. This represents the longest duration investigation of complex rotators to date.

The morphology of all of the phase folded light curves for these objects appear differently in Cycle 3 when compared to those from Cycle 1. Note that we are not claiming that the phase position of features in Cycle 3 corresponds to those in Cycle 1. Due to the rapid nature of the rotation and the almost two years in between Cycle 1 and Cycle 3 observations, we would need to be confident in a rotation period precise to the order of seconds to be certain the phase folded position is lined up to within 0.1 day. This is unreasonable as the sensitivity of period detection is roughly on the order of minutes, and we currently don't have a physical model for the phenomenon that motivates a stability on this time scale. While it is tempting to try to match features from one Cycle to the next and claim they are the same, we cannot be certain that they are directly related. Therefore we plot the light curves with a reasonable period and simply compare the general shape of the two epochs of phase folded light curves holistically. Direct feature to feature comparison is not the intention of this work.

All of the phase folded light curve shapes show significant evolution from Cycle 1 to Cycle 3, to the point of being unrecognizable. We quickly comment on each object, but in general they all maintain similar ranges in amplitude and have sudden brightening/dimming features ($<$ 0.1 day) in both Cycle 1 and Cycle 3. The period of rotation is the same within reasonable uncertainties. 

TIC 38820496 in Cycle 1 had essentially a single small dip and a relatively flat light curve otherwise. In Cycle 3 there is also a single dip that seems to be non existent by the second orbit in sector 28. There is also a larger period of overall increase in brightness in Cycle 3 (starting at around 0.35 day in phase).

TIC 201789285, TIC 234295610, TIC 425933644, and TIC 425937691 all maintain a similar range in their brightness, but it is difficult to identify any features that are the same from Cycle 1 and Cycle 3.

TIC 206544316 has a relative period of flatness that is roughly the same length in both Cycle 1 and Cycle 3 (approximately 0.15 days long, starting at phase 0.2 day in both Cycle 1 and Cycle 3). While the duration of this relatively stable period of brightness is the same, the features outside of this time frame are not alike between the two cycles.

\begin{figure}
\begin{center}
\includegraphics[width=0.5\linewidth]{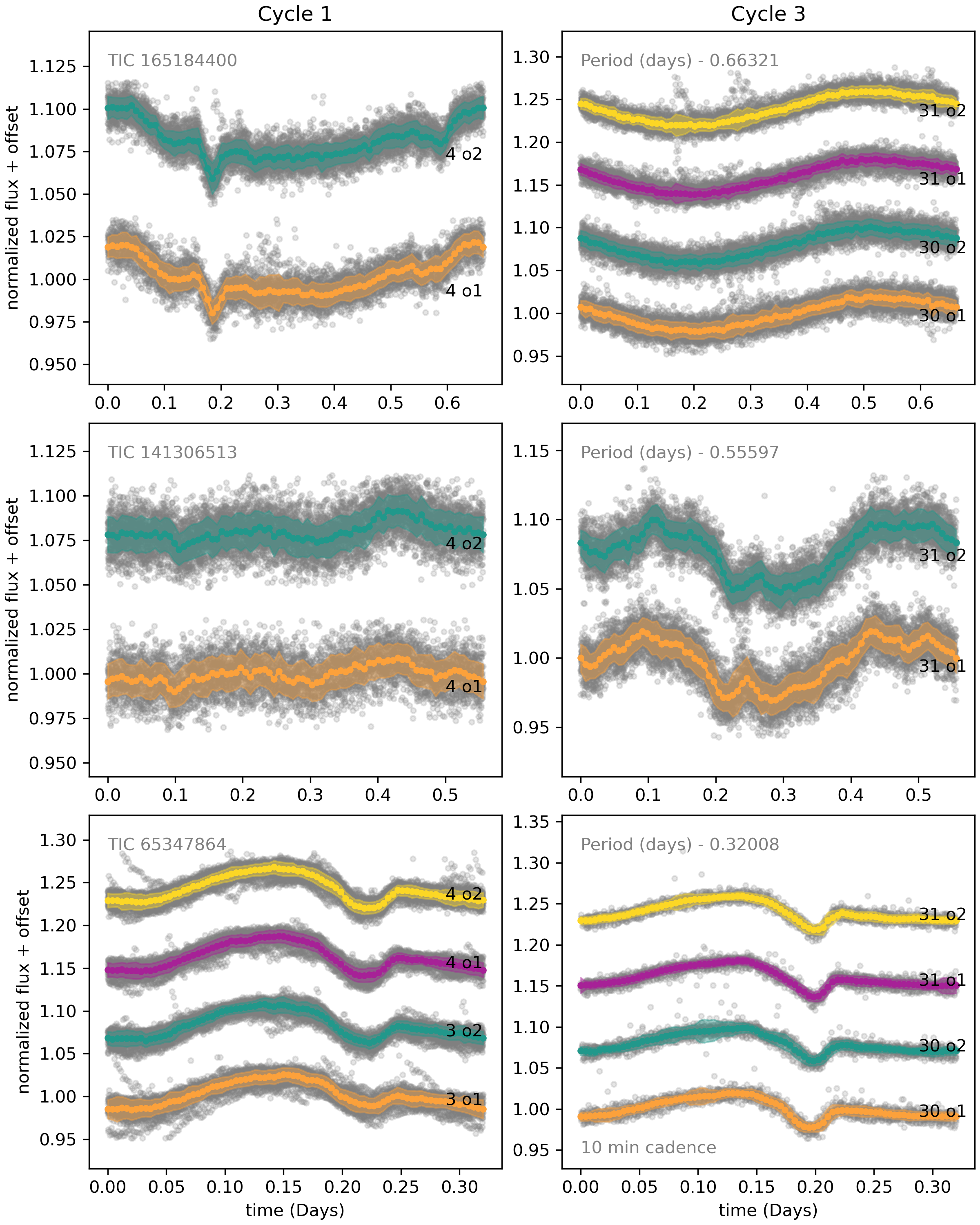}
\caption{Same as in Figure~\ref{fig:scallop_prev}, but now for newly identified complex rotators TIC 165184400, TIC 141306513 and TIC 65347864. Just as in Figure~\ref{fig:scallop_prev}, we note a change in features for all objects from Cycle 1 to Cycle 3, which constrains the stability of this phenomenon to timescales less than a year. Furthermore, we notice a complete disappearance in Cycle 3 of the sharp features in Cycle 1 for TIC 165184400.}
\label{fig:scallop_new}
\end{center}
\end{figure}

In Figure~\ref{fig:scallop_new} we present the newly identified complex rotators from this work. They are presented in the same format as the previously identified objects in Figure~\ref{fig:scallop_prev}. There are important distinctions to be made for each object between Cycle 1 and Cycle 3. For TIC 141306513 The light curve morphology is different between the two Cycles, as well as having approximately double the amplitude in Cycle 3. 

TIC 165184400 (upper right panel of Figure ~\ref{fig:scallop_new}) shows a dramatic difference in the light curves from Cycle 1 to Cycle 3. Sector 4 shows the characteristic sudden dips of a complex rotator, which are extremely stable across the sector. While Sector 4 is the only one with two minute cadence in Cycle 1, we note that we were also able to produce a Sector 3 light curve from the full frame images, which also showed similar features. However, in Sectors 30 and 31 the light curve does not appear like a complex rotator at all. There is a steady increase and decrease of brightness in a simple sinusoidal pattern.

We believe this to be due to the object, and not some form of contamination. While there is a relatively close nearby star of comparable magnitude, if additional flux from a nearby star is to blame for the loss of features it would require the other star to be rotating at the exact same rotation rate, as the period of modulation is the same as that in Cycle 1. This change from complex rotator to simple rotator has not yet been described in the literature.

Finally, TIC 65347864 is the only complex rotator that has the same morphology in both Cycle 1 and Cycle 3. Only FFIs are available for Cycle 3, which we note in the Figure. The overall shape of the phase folded light curve is nearly indistinguishable across sectors as well as between the two Cycles. 

\subsubsection{Complex rotators discussion}\label{complex_discuss}

Previous work on complex rotators \citep[e.g.][]{Stauffer2020_scallop_lcen_crux_ucen_lup,Zhan2019_scallops_tess} commented on the stability of the intricate light curve morphology of these objects over months or, in the case of TIC 177309964 in \citep{Gunther2020_scallop_spec}, over $\sim$1 year in the TESS continuous viewing zone. While this stability is also true for the consecutive sectors in Cycle 1 and Cycle 3 in our objects, our work shows that the long-term evolution in light curve shape in complex rotators is common occurring in 8/9 of our targets over the course of 2 years. 

Furthermore, this evidence of changing features weakens the claim from \citet{Gunther2020_scallop_spec} that the complex rotator features are color dependent. That work took ground based followup data from the SPECULOOS Southern Observatory (SSO) approximately a year after TESS Cycle 1 (Figure 10 in \citealt{Gunther2020_scallop_spec}). They noticed a change in the features at different wavelengths for four objects, including three from our sample (TIC 201789285, TIC 206544316, TIC 425933644), compared to the TESS Cycle 1 features. As shown in Figures~\ref{fig:scallop_prev}--\ref{fig:scallop_new}, the features change dramatically over the course of $\approx$two years so it is not clear if it is appropriate to compare the SSO bandpass light curves to those in TESS Cycle 1 as they were a year removed. \cite{Gunther2020_scallop_spec} noticed different amplitudes and changes in minor features across the SSO bandpasses compared to the TESS light curve. However, given our result that features change dramatically (Cycle 1 and Cycle 3 light curves are almost unrecognizable from one another) over the timescale of years for our objects, truly synchronous observations will more clearly define the extent of color dependence on the features. This work therefore highlights the need for simultaneous monitoring to accurately compare color dependence. 

We end this section by commenting on the likelihood of the potential mechanism of these complex rotators. \cite{Gunther2020_scallop_spec} concludes that the two most likely explanations are either co-rotating clouds of material at a Keplerian orbit or spots and a spin-orbit-misaligned dust disk. Firstly, if all complex rotators are caused by the same phenomena, then the stability of the light curves is certain only on the order of weeks to months, and not years. As seen in Figures ~\ref{fig:scallop_prev}--\ref{fig:scallop_new} the light curves are almost all unrecognizable between Cycles 1 and 3. This timeframe of stability could be incorporated into both theories, the drift of co-rotating clouds could describe the change in the features, and star spot evolution is expected on this time revealed by the misaligned disk. 

Secondly, the disappearance of features in TIC 165184400 is perhaps the most compelling addition to our understanding of complex rotators. A simple sinusoid light curve as seen in Cycle 3 can be interpreted as spot modulation, be it an amalgamation of spots or a few large ones. However the vanishing of the complex rotator features implies the mechanism behind the anomalous light curve is fleeting or transient. This does not necessarily favor the clouds or misaligned disk theory. It is worth noting that other examples in the literature of rapid transient phenomena such as in dippers \citep[e.g.][]{cody_2014_dippers,kesseli_2016_model_dip} rely on a circumstellar disk. The long term variability in the T Tauri system KH 15D is explained by a change in the geometry of the system due to precession \citep{windemuth_2014_kh15d}. While we cite examples that feature disks, we do not favor one theory over the other but these observations do add a new complication to the discussion.

The necessary gaps in the TESS observing coverage of these objects means we cannot describe how the features evolve over the yearlong timescale for Tuc-Hor objects. Investigating the evolution of complex rotator TIC 177309964 \citep{Zhan2019_scallops_tess} that is in the Cycle 1 and Cycle 3 continuing viewing zones would give valuable information about this evolution. Furthermore, considering that a single rotation cycle can be observed over the course of a night, and that small features are mostly stable over the course of a month, it should be possible to track changes over the course of a year with just one night a month from ground based observatories. 

\subsection{Candidate Binaries}\label{binaries}

During the visual inspection of light curves we flagged the presence of multiple periods in the periodograms that are not harmonics of one another. Unresolved binary systems can produce signatures like these \citep[e.g.][]{stauffer_binarys,tokovinin_2018}, as star spots on each star contribute a different periodic signal to the light curve. This is potentially exacerbated by the large pixel size of TESS. If multiple periods are sufficiently different, they are visible as distinct patterns in the light curve and distinct peaks in the periodogram. In cases where the periods are close to one another what is often seen is a beat pattern in the light curve \citep[e.g.][]{paudel2019}. A beat pattern can be caused by two stars each with their own star spots and slightly different rotation periods, but could also plausibly be caused by two star spots on the surface of a single star but at different latitudes experiencing differential rotation. 

The RUWE statistic produced through the Gaia catalog represents another potential sign of an unresolved multiple object system. If the single-star astrometric solution \citep{Lindegren_RUWE,Lindegren_RUWE_edr3} struggles to fit the object's position time series it could be due to an unresolved companion \citep[e.g.,][]{belokurov_ruwe,ruwe_2021}. However, \citet{Palumbo_2021} showed that there is a color dependence with RUWE in Tuc-Hor objects, with redder objects having slightly higher RUWE values. While this doesn't disqualify RUWE from being a useful metric for probing potential binarity in Tuc-Hor, it does encourage any conclusions to be made with caution for these fainter objects.

In Table~\ref{table:binary} we present 19 objects for which we identified multiple periods in their TESS light curves. Along with the TIC and Gaia DR2 ID we also include the two distinct periods measured along with spectral type where available, a flag for evidence of beating in their light curve, the TIC contamination ratio \citep{TIC} and Gaia DR2 RUWE \citep{gaiadr2}. Objects that have been described as binaries in the literature have a source of the citation in the Binary source column.

\begin{deluxetable*}{rhrrrrrrrrr}
\tablecaption{Multi Period Light Curves in Tuc-Hor.\label{table:binary}}

\tablehead{
\colhead{TIC ID} &  \nocolhead{Gaia DR2 ID} & \colhead{RA} & \colhead{Dec} & \colhead{SPT} & \colhead{Period 1} & \colhead{Period 2} & \colhead{Beat Flag} & \colhead{TIC Cont Ratio}  & \colhead{Gaia DR2 $RUWE$} & \colhead{Binary Source}}

\startdata
201898222 & 4741424237273655040 &  40.509294 & -53.987466 &   M4.0 &  0.445723 &   0.56000 &          1 &                 0.774935 &    1.141236 &             -\tablenotemark{\footnotesize a}\tablenotemark{\footnotesize 1} \\
201898220 & 4741424615230777216 &  40.517656 & -53.983394 &   M3.5 &  0.568254 &   0.44500 &          1 &                 0.330944 &    2.390334 &             -\tablenotemark{\footnotesize 1} \\
166852312 & 4742040410461492096 &  40.445841 & -52.997949 &   K6 V &  0.519252 &   0.55900 &          1 &                 0.291826 &    1.918205 &          \tablenotemark{\footnotesize a}\tablenotemark{\footnotesize 2} \\
166852313 & 4742040513540707072 &  40.447775 & -52.991896 & M2.5 V &  0.559470 &   0.51900 &          1 &                 2.217621 &    2.133387 &           S17\tablenotemark{\footnotesize 2} \\
165941376 & 4944355154177076864 &  31.757933 & -44.110697 &   M1.0 &  6.582363 &   0.39000 &          0 &                 0.096210 &    1.132751 &             X\tablenotemark{\footnotesize 3} \\
117874958 & 5000558409016727296 &   9.897819 & -38.288525 &   (M4) &  6.214116 &   1.04900 &          0 &                 2.683233 &    1.038216 &             -\tablenotemark{\footnotesize 4} \\
\hline
\hline
50311775 & 4638630860832778240 &  30.730780 & -72.987675 &    NaN &  0.232935 &   0.19300 &          1 &                 0.007163 &    3.044104 &             - \\
631284388 & 4696427857777404544 &  34.541066 & -66.964644 &   M4.5 &  0.343798 &   0.31900 &          1 &                      NaN &         NaN &           \citet{Janson_binary_2017} \\
631547535 & 4739713087942439168 &  37.441135 & -55.697190 &   M4.0 &  0.490647 &   0.79000 &          0 &                      NaN &         NaN &           \citet{shan_binary_2017} \\
231277551 & 4741900291448836736 &  42.128317 & -52.994430 &    NaN &  0.217231 &   0.14000 &          0 &                 0.047809 &    1.761249 &             - \\
231294802 & 4748105247880681216 &  44.829988 & -51.376189 &   M5.0 &  0.204511 &   0.46810 &          0 &                 0.000065 &    1.232498 &             \tablenotemark{\footnotesize a} \\
350712873 & 4764908637411876736 &  88.204184 & -59.485424 &   (M5) &  0.122113 &   0.12740 &          1 &                 0.008479 &    1.304503 &             - \\
685945639 & 4843959091042177408 &  61.415438 & -40.236237 &   M4.2 &  0.689013 &   0.44410 &          1 &                      NaN &   11.909748 &           \citet{shan_binary_2017} \\
44670258 & 4888246251178417024 &  60.016327 & -29.037955 &   K4.6 &  4.574736 &   0.31000 &          0 &                 0.721228 &    3.772712 &          \citet{wide_binary_andrews_2017} \\
150068381 & 5477559813377591552 &  91.871983 & -64.158836 &   (M4) &  0.534128 &   0.17642 &          0 &                 0.051127 &    1.305567 &             - \\
738102496 & 5571239441011080960 &  90.095852 & -44.022580 &   M4 V &  0.913970 &   0.59368 &          1 &                      NaN &    1.265127 &             - \\
2054862814 & 6388075665896356480 & 350.570927 & -68.550402 &    NaN &  2.349246 &   0.40000 &          0 &                      NaN &    1.821980 &             - \\
344552743 & 6399670771926369152 & 321.960975 & -68.684622 &   M4.0 &  0.342975 &   0.20000 &          0 &                 0.005328 &    1.469350 &             - \\
238813187 & 6402514555672235264 & 327.271222 & -64.218181 & M4.5 V &  0.168912 &   0.17558 &          1 &                 0.006295 &    2.408733 &           \citet{shan_binary_2017} \\
\enddata
\tablenotetext{a}{was seen as single in \citet{Janson_binary_2017}}
\tablenotetext{1}{TIC 201898222 and TIC 201898220 are likely effecting each others light curves.}
\tablenotetext{2}{TIC 166852312 and TIC 166852313 are likely effecting each others light curves.}
\tablenotetext{3}{TIC 165941376 is likely effecting this object's light curve.}
\tablenotetext{4}{TIC 117874959 is likely effecting this object's light curve.}
\tablecomments{A list of objects for which we found there were multiple, non-harmonic periods in the light curves. Objects above the double line were noted to have a nearby Tuc-Hor member in Section ~\ref{contamination}, and the multiple periods for these objects are assumed to be due to the nearby Tuc-Hor star.}
\end{deluxetable*}

Table ~\ref{table:binary} is divided into objects that were noted in our contamination analysis in Section ~\ref{contamination} and those that were not. For the six systems in the first part of the table we assume that the second rotation period we measure is due to the other Tuc-Hor member. The other 13 systems in the lower part of the table do not have a known nearby Tuc-Hor member. Of the multi-period objects across both parts of the table, 11 out of 19 have no evidence of binarity in the literature. From those 11, TIC 117874958 has a TIC Cont Ratio $>$ 1, with three Gaia DR2 sources within 20$''$. This suggests that the additional signal could be coming from a nearby object.

\begin{figure*}
\begin{center}
\includegraphics[width=\linewidth]{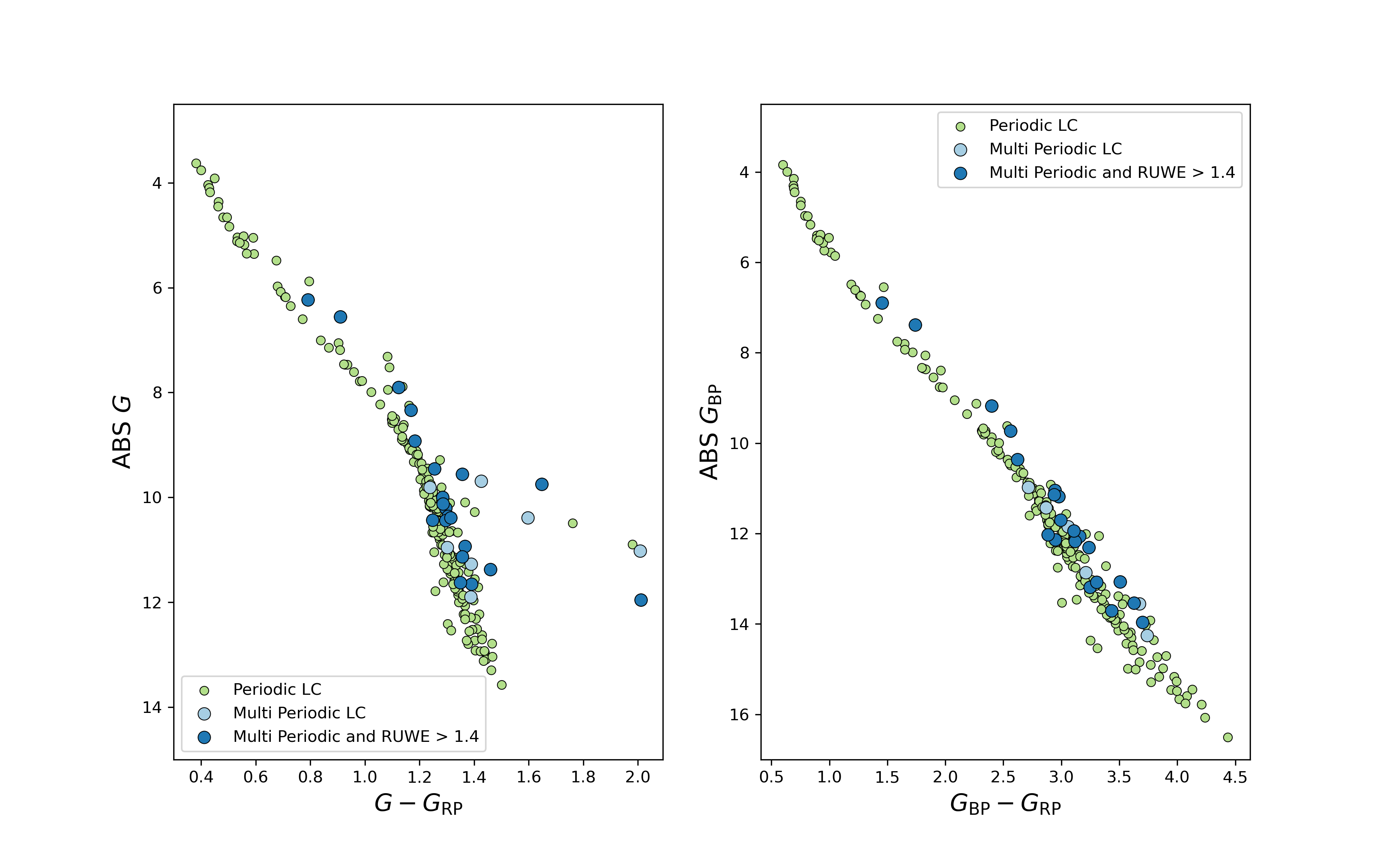}
\caption{Two Color-Magnitude Diagrams of the objects for which we recovered rotation periods. Those with multiple periods in their light curves, as well as those with multiple periods and high Gaia DR2 RUWE values are highlighted in light blue and dark blue respectively. The two different Gaia colors present slightly different main sequences, and the multi periodic objects are mostly part of a binary sequence. The 6 objects that stand out as redder than \grp{} $=$ 1.6 in the left panel are likely due to spurious G magnitudes.}
\label{fig:cmd_binary}
\end{center}
\end{figure*}

In Figure ~\ref{fig:cmd_binary} we present a color-magnitude diagram (CMD) of the objects with rotation periods measured from light curves in Tuc-Hor. Sources that are over-luminous and above the main sequence on the CMD are likely to be binaries. We highlight those with multiple periods as light blue points, and those with multiple periods and a Gaia DR 2 RUWE $>$ 1.4 as dark blue. We present both the \grp{} and $(G_{\rm BP} - G_{\rm RP})$ CMD's, as they highlight the binary sequence better in different regimes.


From Figure ~\ref{fig:cmd_binary} we see that multiple rotation periods can be a useful way to identify potential binary systems. It is not a completely successful method, as there are many objects  on the binary sequence which do not exhibit multiple periods. The inverse is true as well, where there are objects that appear to be regular main sequence objects that contain multiple periods. However, it does provide compelling evidence for candidate binary objects. This is especially true among the M dwarfs where the binary sequence isn't as clear \citep{stauffer_binarys,tokovinin_2018}. We consider it a useful indicator for potential follow up observations. The 6 objects that are redder than \grp{} $=$ 1.6 in the left panel are likely due to spurious G magnitudes, as they converge with other objects in the $(G_{\rm BP} - G_{\rm RP})$ panel. We limit the upper range of further plots to \grp{} $=$ 1.6 to focus our conclusions on the majority of the objects.

\section{Rotation Period Distribution of Tuc-Hor}\label{sec:gyro}

From our measured rotation periods we can describe the rotation period distribution for Tuc-Hor. Color--period plots carry information about the age of a population of stars based on how slowly some objects are spinning. In particular, the slow rotator sequence is an important feature, as it represents stars that are evolving most similarly to ``Skumanich like'' power law relations \citep{skumanich_1972}. There is a mass dependence as to how long it takes a star to converge on to this sequence, with higher mass stars converging earlier.

\begin{figure*}
\begin{center}
\includegraphics[width=\linewidth]{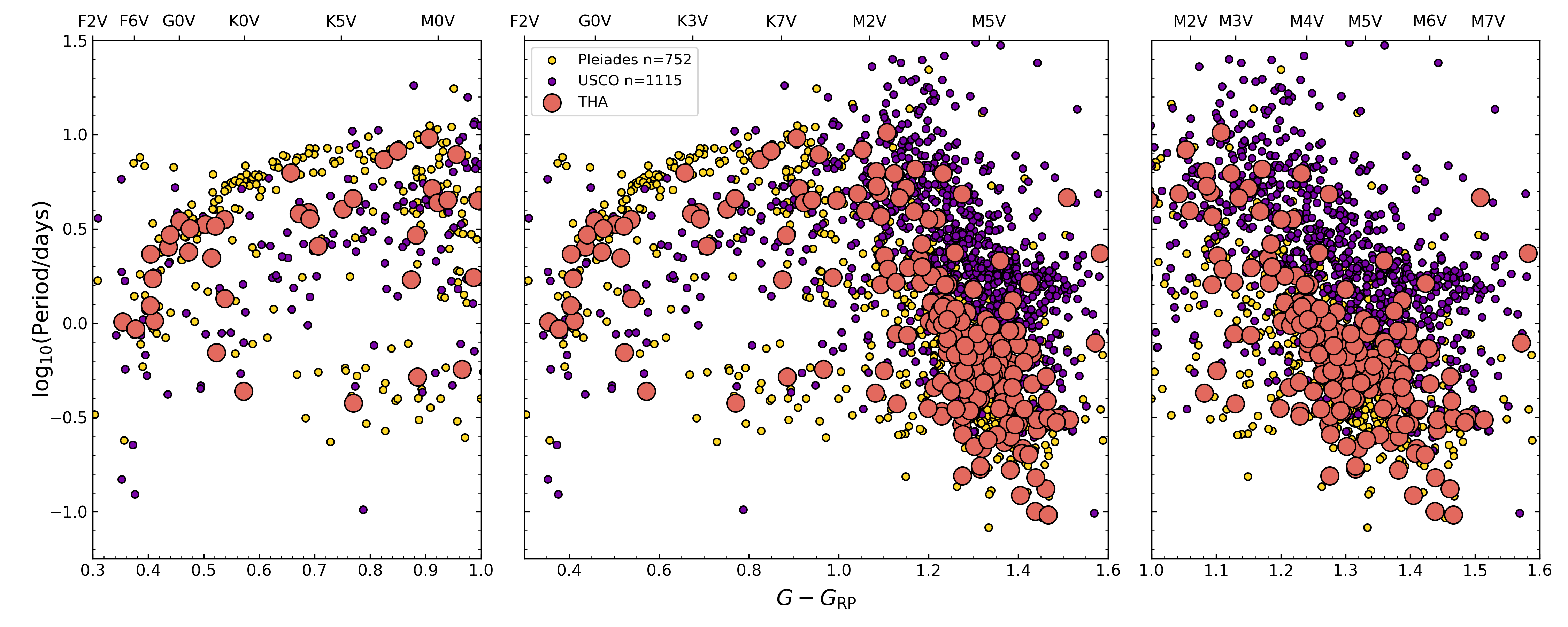}
\caption{A comparison of the rotation period - color distribution of Tuc-Hor against USCO and Pleiades. The left panel shows the comparison between associations at \grp{} $<$ 1.0, which focuses on the slow rotator regime at this age. We find that F and G spectral types are converged, but it is challenging to tell for K stars, in part due to lack of members at these spectral types in our sample. The right panel shows \grp{} $>$ 1.0 which represents the M dwarf regime. The M stars agree far more with the Pleiades, and are spun up when compared to those in USCO. The middle panel encompasses the full rotation period distribution presented in this work.}
\label{fig:color_rot_comp}
\end{center}
\end{figure*}

In Figure ~\ref{fig:color_rot_comp} we show all measured rotations rates on a color--period plot, split into 3 views, while placing Tuc-Hor in comparison to Upper Scorpius (USCO) at $\sim$ 5 Myr \citep{Rebull2018_usco} and Pleiades at $\sim$ 120 Myr \citep{Rebull2016_plei}. The middle panel shows the full rotation period distribution of the association. The left panel focuses on the higher mass and bluer objects with \grp{} $=$ 0.4--0.6 that are beginning to converge onto the slow rotator sequence. The right panel focuses on redder colors, where the slow rotator sequence is essentially undefined but there is a broad distribution of rotation periods from 10 days to $<$1 day across the range of \grp = 1.0--1.5, with a shortening in rotation period for redder objects. Rough spectral types that correspond to a given \grp{} color run across the top of the plot. \footnote{based on \url{ http://www.pas.rochester.edu/~emamajek/EEM_dwarf_UBVIJHK_colors_Teff.txt}} We discuss each panel further below.

From the middle panel in Figure~\ref{fig:color_rot_comp} we confirm that the rotation period distribution of the Tuc-Hor objects is consistent with the larger trends seen in other young associations. We compare the rotation rates measured in this work to those of Upper Scorpius (USCO) and Pleiades, associations that bracket the conventional age of Tuc-Hor. We use the rotation periods for USCO and the Pleiades from \citet{Rebull2018_usco} and \citet{Rebull2016_plei} respectively. Both of these works made use of light curves from the K2 mission, and have a $\sim$ 90 day baseline from a K2 campaign. This means that they are sensitive to a longer period than a single TESS sector. However as we show in Section~\ref{data}, the median number of TESS sectors for our objects is close to four, which should be  sensitive to a period of at least $\sim$ 24 days. We therefore find it reasonable to compare our rotation rates to these data sets.

Looking at the higher mass stars in the distribution in the left panel of Figure ~\ref{fig:color_rot_comp}, compared against the younger USCO, the slow rotator sequence between \grp = 0.4-0.6 in Tuc-Hor is not present in USCO. There are some objects at similar rotation periods in USCO, but the distribution is not well constrained. In comparison to the Pleiades, Tuc-Hor is remarkably similar up to \grp{}$\sim$0.6. This boundary might be an effect from our target selection and membership lists, as we have relatively few objects with \grp{} colors between 0.6--0.85. However, for this same color range in Pleiades, they tend to be constrained to the slow rotator sequence. In contrast, only $\sim$ 1/8 of Tuc-Hor members in this color range has reached the slow rotation sequence.

Finally, focusing now on the right panel of Figure~\ref{fig:color_rot_comp} and the range of \grp{} $>$ 1.0 we turn to the M dwarf regime. \citet{popinchalk_2021} contains a compilation of M dwarf rotation rates, and a thorough discussion of the angular momentum evolution across age, especially in young groups. Starting with a comparison with USCO, we see that the USCO objects are generally more slowly rotating when compared to Tuc-Hor objects of the same color. The spread in the rotation distribution is greater in USCO as well, even though the general trend of shorter rotation periods for redder colors is present in both. When compared to the Pleiades, the Tuc-Hor distribution is far more similar. The small exception is at \grp{} $\simeq$ 1.0$\sim$1.2 where there appears to be an additional population of objects with $log_{\rm 10}(P_{\rm rot})$ $>$ 0.5 when comparing Tuc-Hor to the Pleiades. The spin up in M dwarfs in USCO age to Tuc-Hor age is thought to be due to the star formation process with the objects still condensing and collapsing over tens of million of years. If this over-density of longer periods in \grp{} 1.0-1.2 is real in Tuc-Hor compared to the Pleiades, it could mean that the spin down in these objects is slower compared to redder colors.

At 40 Myr, there is a chance that some of the later M's are brown dwarfs and will eventually cool. Identifying those objects is beyond the scope of this paper, but we refer the reader to \citet{faherty_2016} for a list of young brown dwarfs, and \citet{2022_vos_bdwarfspin} for the most up to date census of rotation periods in young brown dwarfs.



\section{Diagnosing the sample: beyond light curves}\label{sec:diag_2}
\subsection{IR Excess}\label{sec:IR}

We investigate the objects in our Tuc-Hor Sample for evidence of disks or dust in the form of Infrared (IR) Excess. \citet{Rebull2018_usco} showed that 18\% of objects within Upper Scorpius with infrared excess had a noticeably different rotation period distribution compared to those without. This infrared excess was attributed to being caused by disks around objects within the 10 Myr group. In particular in the $\sim$ M dwarf regime there was a pile up at P(rot) $\approx$ 2 days. No such infrared excess was found in the Pleiades \citep{Rebull2016_plei} for the M dwarfs, as there was only evidence of debris disks in a few G type stars. Tuc-Hor is therefore placed at a potential transition point between disk-bearing groups at ages of $<$100 Myr and the disk-free groups at ages $>$100 Myr.

\citet{kraus_2014_tuc_ha_li} searched for IR excess by using WISE \citep{wise_2010} photometry for their Tuc-Hor membership list and found no evidence. \citet{ldisk_boucher} did find IR excess for a candidate L dwarf in Tuc-Hor. We used the same methodology in \citet{kraus_2014_tuc_ha_li} and \citet{Rebull2016_plei} of comparing W1 and W3.  We draw the same conclusion as \citet{kraus_2014_tuc_ha_li} for our extended membership list in that there is not evidence for IR Excess in any Tuc-Hor object. Furthermore there is no noticeable build up of rotation periods in the low mass regime. We discuss the implications of this result in Section ~\ref{sec:dis:rot}.

\subsection{H$\alpha$ and Lithium Equivalent Widths}\label{halpha_li}

\begin{figure}
\begin{center}
\includegraphics[width=0.75\linewidth]{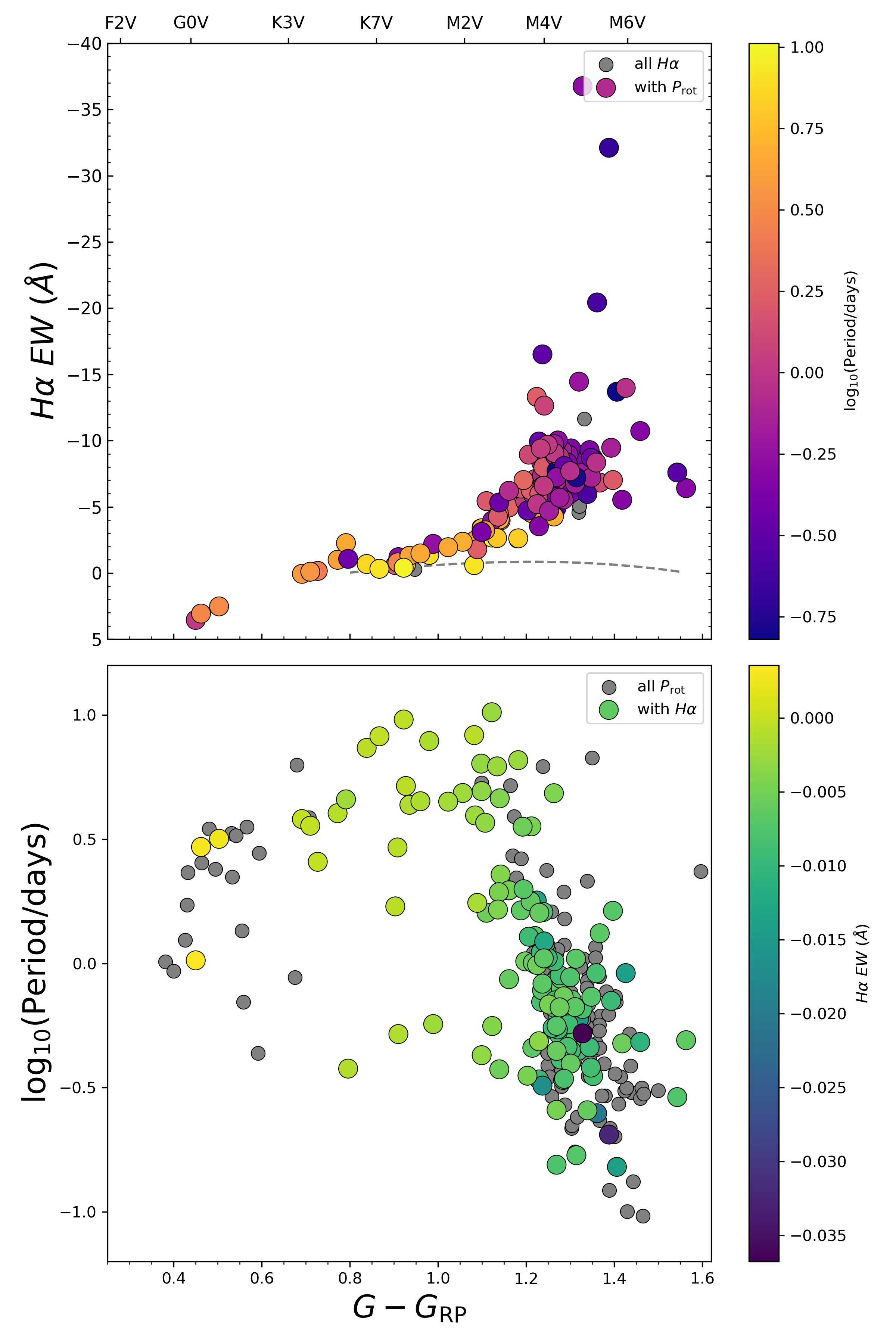}
\caption{The relationship between \ha{} and log $P_{\rm rot}$ across \grp{} color. \textit{Top} \ha{} equivalent width against \grp{}, color-coded by log $P_{\rm rot}$. Objects within our membership list that have an \ha{} measurement but no rotation are plotted as grey circles. \textit{Bottom} The rotation period distribution with \grp{} color-coded by \ha{} equivalent width. \ha{} values are generally independent from rotation period.}
\label{fig:halpha}
\end{center}
\end{figure}

We plot the stars in a color scale related to their rotation period in Figure~\ref{fig:halpha}. The equivalent width of the \ha{} emission line for Tuc-Hor members increases with decreasing mass. This is due to changes in photospheric contribution and not necessarily more magnetic activity \citep{stauffer_hartmann_1986,kiman_2021_ha}. We include the M dwarf activity boundary from \citet{kiman_2021_ha} and see that almost all objects in the M dwarf regime ($\sim$ \grp{} $\geq$ 0.9) are active, which is consistent with their young age \citep{popinchalk_2021}. 

For 0.6 $<$ \grp{} $<$ 1.2 we notice that the \ha{} relation is relatively tight despite the large range of rotation periods of the objects at a given color. There are relatively few objects at 0.6$<$ \grp{} $<$ 1.0, but the relation holds through to at least \grp{}$=$ 1.2. This could signal that at this age, \ha{} values are independent from rotation period (See Section~\ref{discuss}). For \grp{} $>$ 1.2 we have our largest density of objects, and we note perhaps a slight trend of increasing \ha{} EW with shorter rotation period and redder \grp{} color. This is expected as both short rotation period and large \ha{} EW are expected to be related to magnetic activity. \citep[e.g.][]{skumanich_1972,mamajek_hillenbrand,west_2015_activity_rotation}

\begin{figure}
\begin{center}
\includegraphics[width=0.75\linewidth]{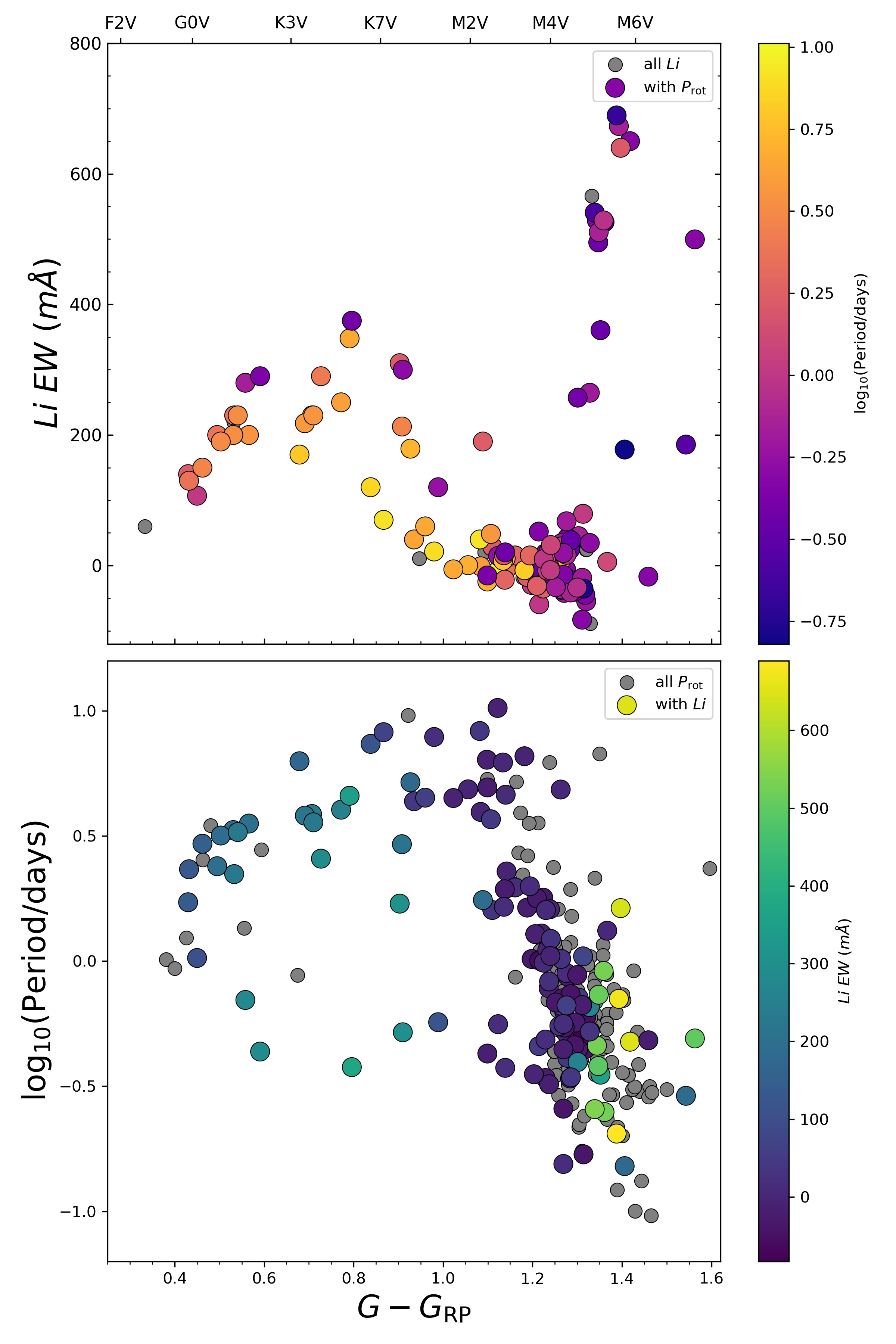}
\caption{The same as Figure \ref{fig:halpha} but with Li equivalent width instead of \ha{}. Objects within the color range \grp{} $<$ 1.0 that have slower rotation periods also have the smallest equivalent widths.}
\label{fig:lithium}
\end{center}
\end{figure}

In Figure~\ref{fig:lithium} the equivalent width of the \ion{Li}{1} 6708 \AA{} line is plotted against \grp. \citet{kraus_2014_tuc_ha_li} defined a depletion boundary based on spectral types for Tuc-Hor. \grp{} is a proxy for spectral type, but we are able to define the lithium depletion boundary again in terms of this new variable at \grp{}$\sim$1.0.

When we color code the lithium equivalent width plot by rotation period in the lower panel of Figure~\ref{fig:lithium}, we see that within the color range \grp{} $<$ 1.0, objects with faster rotations have larger lithium equivalent width. Specifically in the color range 0.8 $<$ \grp $<$ 1.0, the object with the the slowest rotation period has the smallest lithium equivalent width. Finally, we note that for objects with \grp $>$ 1.2 there is no conclusive relation between the wide range of lithium equivalent width values and the rotation period.

\subsection{X-ray and NUV luminosity}\label{xray_uv}

\begin{figure}
\begin{center}
\includegraphics[width=0.75\linewidth]{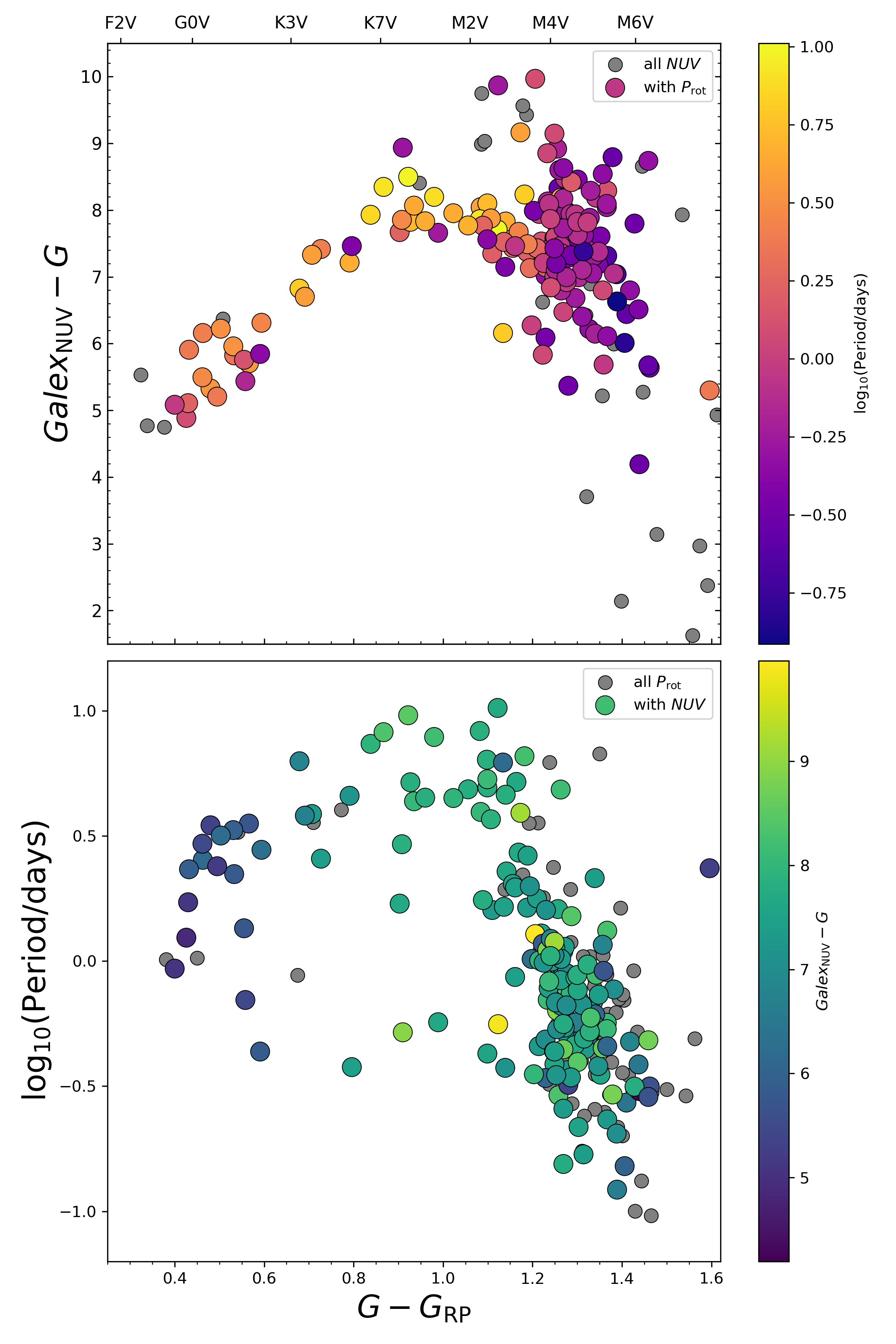}
\caption{The same as Figure \ref{fig:halpha} but with $Galex_{\rm NUV}-G$ instead of \ha{}. There is a tight relation in $NUV-G$ for 0.4 $<$\grp{} $<$ 1.2, there is no clear rotation dependence. For  \grp{} $>$ 1.2 there is a wider range of $NUV-G$ values that spans three magnitudes, but still shows no rotation period dependence.}
\label{fig:nuv}
\end{center}
\end{figure}

We present the relation between ($NUV-G$) in Figure~\ref{fig:nuv}.
Similarly to the \ha{} equivalent width, the ($NUV-G$) relation is quite tight in 0.4 $<$\grp{} $<$ 1.2 with ($NUV-G$) increasing with color up to a peak at \grp{} $=$ 1.0 before starting to decrease again. After \grp{} $=$ 1.2 there is a larger distribution of ($NUV-G$) values, which is consistent with other young associations \citep[e.g.][]{gagne_mutau_2020}. Also similar to \ha{}, within the tight relation in the range of 0.4 $<$ \grp{} $<$ 1.2, there does not appear to be a dependence on rotation period. Within a \grp{} color in this range the ($NUV-G$) values are the same regardless of rotation period. We also note there are a few objects at \grp{} $=$ 1.2 that have unusually high ($NUV-G$) $>$ 9 and seem to sit above the rest of the objects. 

\begin{figure}
\begin{center}
\includegraphics[width=0.75\linewidth]{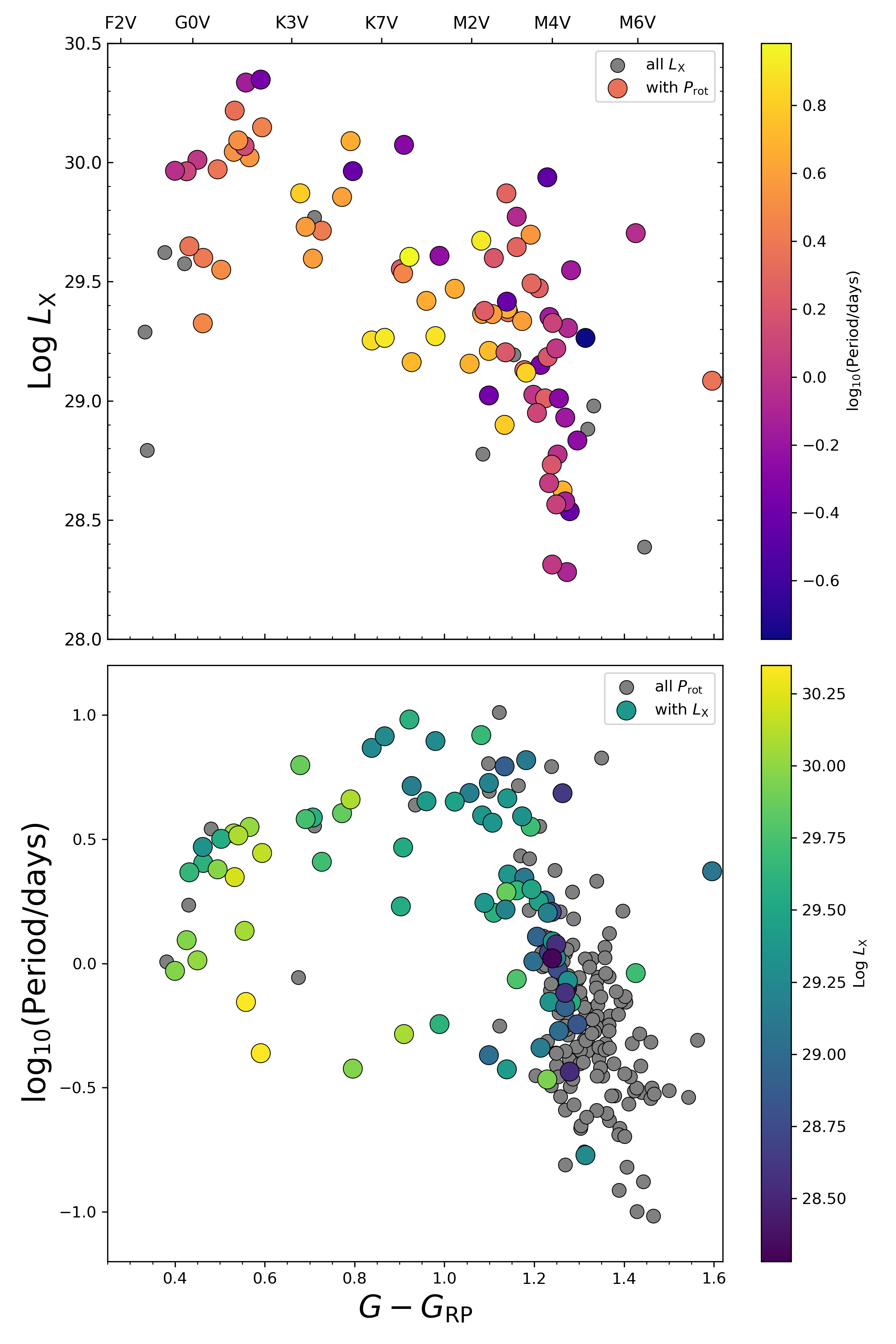}
\caption{The same as Figure \ref{fig:halpha} but with Log $L_{\rm X}$ instead of \ha{}. There is little trend between rotation period and $L_{\rm X}$, as for a given color there exists a range of rotation periods and X-ray luminosity.}
\label{fig:xray}
\end{center}
\end{figure}

Finally, we present the log of X-ray luminosity in Figure~\ref{fig:xray}. We convert the fluxes from ROSAT \citep{ROSAT_2016} listed in Table \ref{table:sources} to magnitudes using the relation 
\begin{equation}
    L_{\rm X} = 1.20*10^{38} * f_{\rm X} * D^2
\end{equation}
where $f_{\rm X}$ is the X-ray flux drawn from hardness ratio 1 \citep{ROSAT_2016}, and D is the distance to the object for which we use Gaia DR2 parallaxes to calculate.

The X-ray luminosity versus \grp~relation fails to extend as deeply into the M dwarfs as the other indicators of youth. This is in large part due to the lack of X-ray detections, as the smaller radii of the low mass stars make them intrinsically less luminous.

We note that in general there is little trend between rotation period and $L_{\rm X}$, and that at a given color there exists a range of rotation periods and X-ray luminosity. This range appears to be slightly wider at \grp{} = 1.2, spanning 3 orders of magnitude. This is near the edge of the detection limit, but it does seem to be similar to the range of values at \grp{} = 1.2 noted in the $NUV-G$ in Figure~\ref{fig:nuv}. It is not clear if this broadening of the distribution manifests from a wider range of astrophysical activity in these objects, or is an effect of the difficulty of observing these smaller objects.

\subsection{Objects with evidence of youth but no measurable rotation period}

We end by explicitly drawing attention to the fact that for each of the indicators of youth presented, there were some objects that we did not report a rotation period for despite the indication of youth. For example, in the case of \ion{Li}{1} equivalent width measurements, there were 11 objects where this was the case. Of those 11 objects, 6 were not viewed by TESS, 4 showed some signs of variability but ultimately were not accepted as true periods, and 1 was deemed flat.

We start from an assumption that all young stars are active enough to have appreciable star spot induced modulation. A lack of periodic signals for objects that otherwise appear young from other indicators are potentially interesting objects. Potential explanations include the inclination angle of the star being pole on and minimizing the effects of star spot variation, or a stellar minimum in activity and star spot coverage similar to the Sun's 11 year cycle.

\section{Discussion}\label{discuss}

\subsection{Rotation Period as an Indicator of Youth}\label{dis:rot_youth}
It is important to appreciate the power of rotation rates as a tool of confirming an object's youth. As was shown in Sections~\ref{halpha_li} and \ref{xray_uv}, most objects that had previous identifications of youth were found to have a rotation period in this work. This shows that rotation period truly is another complementary tool for identifying and confirming young objects at this age. (The lack of dusty disks that add stochastic components to light curves makes it more straightforward to identify rotation periods)

Furthermore, it is perhaps the most valuable among them when considering extended associations of young stars as rotation periods are quickly becoming the easiest signatures of youth to obtain. The TESS mission provides nearly all sky coverage, and with a single sector is sensitive to rotation periods of $\lesssim$10 days. As the mission continues, the entire sky will eventually be covered. For nearby loose populations of stars similar to Tuc-Hor, our work shows that contamination does not seem to be a significant issue even if a small subset of stars need to be considered on a case by case basis (see Section~\ref{contamination}). If rotation rate is as valid an indicator of youth as the others that have previously been established, it points to a future of unprecedented advances in identifying young moving groups. This power combination of the open source data products of these two missions (Gaia and TESS) is capable of defining membership of co-moving groups through kinematics and then confirming their youth via gyrochronology.

That is not to say that rotation necessarily replaces the need for other age activity indicators. An individual object's age based on its rotation rate is always potentially in question. Spin up due to a companion can allow older objects to masquerade as younger. Nevertheless it is a powerful tool for rapidly identifying groups of young objects, and combining all the indications of youth is necessary to understand the complex interplay between these observables.


\subsection{Gyrochronology of Tuc-Hor}\label{sec:dis:rot}

Based on the rotation period distribution presented in Section~\ref{sec:gyro}, we find Tuc-Hor to be consistent with an age of 40 Myr, or between USCO and Pleiades. We attribute this age mostly due to the few F and early G type stars that have converged onto the slow rotator sequence, as the period distribution of the rest of the mass range carries far less age information. However it is important to note that we have an unfortunate gap in our color range very close to where the convergence onto the slow rotator sequence appears to be occurring ($G -  G_{\rm RP} \approx 1.0$). This process of convergence has been proposed to be due to individual stars having a change in their magnetic field structure from higher order complexity to one dominated by a dipole \citep{garraffo_2018_revrev}.

We also put our M dwarf rotation rate distribution in context with those in Upper Scorpius. \citet{Rebull2018_usco} noted a build up of objects rotating at $\sim$ 2 days, all with IR excesses according to their WISE colors. We performed the same analysis by using WISE photometry associated with each object according to the TIC catalog and found that none of our Tuc-Hor objects showed any infrared excess. We also did not see a build up in rotation period in the M dwarf regime at 2 days; as discussed in Section~\ref{sec:gyro}, Tuc-Hor was mostly similar to the Pleiades. This would appear to suggest that the infrared excess (considered to be caused by circumstellar disks) dissipates before 40 million years, if we assume a similar evolution between the two groups. This is consistent with the understanding of disk lifetimes \citep{disk_review}, but what it also suggests is that the effect of disk-locking that causes the build up of rotation periods on the angular momentum is effectively forgotten not just by the Pleiades age of 120 Myr, but much earlier at 40 Myr, if not sooner. It will be important to see if this is true for a larger sample size beyond the relatively few objects in Tuc-Hor, but since M dwarf stars don't seem to ``remember'' the effects of this disk locking even on the order of tens of millions of years, it implies that it is a fleeting phase in the M dwarf angular momentum evolution.

Furthermore, the information in the TESS light curves is rich and a valuable asset when examining young populations. The rotation period is truly just the lowest hanging fruit. For example, within this work we have identified new phenomena among complex rotators in Section ~\ref{complex_comment}. Elsewhere in the literature, \citet{Palumbo_2021} found an exciting stellar surface event in a Tuc-Hor object, \citet{adina_flares_2022} has used flare rates to probe magnetic stability of stellar populations, and \citet{thyme_dustuc_2019} has found planetary systems around one of these very active stars. Developing methods for analyzing the change in morphology across years will surely provide valuable insight into constraining stellar cycles, star spot life times, and differential rotation in even the most seemingly stable variable star. The continued mission lifetime of TESS demands further analysis tools as more and more data across the sky becomes available in the future.

\section{Conclusions}\label{conclusions}

Using light curves derived from TESS FFIs, we measured stellar rotation periods for \rotobs{} stars that are considered members of Tuc-Hor based on astrometry from Gaia. We successfully measured rotation periods for \recovrate{} of our sample with usable TESS data, which is a remarkable recovery rate on par with the K2 survey of the Pleiades \citep{Rebull2016_plei}. We determined that sample contamination was not an issue in recovering rotation periods. In other words, TESS excels at characterizing young and active stars. 

Within this sample we found three new complex rotators and analyzed them along with six previously identified in Tuc-Hor to study their evolution between Cycle 1 and Cycle 3. We present the first long-term investigation of complex rotators in the literature, finding that only one of the nine had the same light curve morphology between the two cycles, and out of the eight that changed, one had no sign of complex rotator morphology by Cycle 3. In addition we identified several potential binaries in our overall sample from multiple periods detected in their light curves.

We compared these rotation periods with USCO and the Pleiades, and combined rotation periods with other indications of youth including IR excess, \ha{} emission and lithium absorption equivalent widths, X-ray luminosity and GALEX~NUV$-G$. While contrasting the rotation periods with these other age indicators, we noticed that rotation period does not have a bearing on the relation between \ha{} or NUV$-G$ at \grp{} colors $<$ 1.2. Furthermore, with every age indicator there were more objects in our sample with rotation periods than with individual age indicators. We conclude that rotation period is a powerful tool for confirming membership of young moving group associations because of its ease of measurement and prevalence (nearing all-sky as TESS advances into Cycles 5 and beyond).


\begin{acknowledgments}

This work is supported by NASA TESS GI grant No.\ 80NSSC21K0792 and No.\ 80NSSC19K1708. This work was supported in part by the National Science Foundation under Grant No. NSF PHY-1748958.
J.L.C.\ is supported by NSF AST-2009840 and NASA TESS GI grant No.\ 80NSSC22K0299 (G04217).  J.F acjknowledges the continuous support of the Heising Simons foundation for this work.


This paper includes data collected by the TESS mission, 
which is funded by the NASA Science Mission directorate.
We obtained these data from the Mikulski Archive for Space Telescopes (MAST). 
STScI is operated by the Association of Universities for Research in Astronomy, Inc., 
under NASA contract NAS5-26555. 
Support for MAST for non-HST data is provided by the NASA Office of Space Science via 
grant NNX09AF08G and by other grants and contracts.

This work has made use of data from the European Space Agency (ESA)
mission Gaia,\footnote{\url{https://www.cosmos.esa.int/gaia}} processed by
the Gaia Data Processing and Analysis Consortium (DPAC).\footnote{\url{https://www.cosmos.esa.int/web/gaia/dpac/consortium}} Funding
for the DPAC has been provided by national institutions, in particular
the institutions participating in the Gaia Multilateral Agreement.
This research also made use of public auxiliary data provided by ESA/Gaia/DPAC/CU5 and prepared by Carine Babusiaux.

%

This research has also made use of NASA's Astrophysics Data System, 
and the VizieR \citep{vizier} and SIMBAD \citep{simbad} databases, 
operated at CDS, Strasbourg, France.

\end{acknowledgments}

%

\vspace{5mm}
\facilities{Gaia, TESS}


\software{  Astropy \citep{Astropy2013, Astropy2018},
            Astroquery \citep{astroquery},
            google.colab, 
            IPython \citep{ipython},
            Matplotlib \citep{matplotlib}, 
            NumPy \citep{Numpy2020}, 
            Photutils \citep{Photutils2016},
            TESScut \citep{Astrocut},
            unpopular \citep{hattori_cpm_2021}
            }





\bibliography{biblio}{}
\bibliographystyle{aasjournal}



\end{document}